%% file: main.tex
\documentclass[12pt]{article}

\usepackage{amssymb,amsmath,amsfonts,eurosym,geometry,ulem,graphicx,caption,color,setspace,sectsty,comment,footmisc,caption,natbib,pdflscape,array,hyperref,subcaption,threeparttable, amsthm, rotating,chngcntr,lscape,subfloat}

\newcolumntype{L}[1]{>{\raggedright\let\newline\\arraybackslash\hspace{0pt}}m{#1}}
\newcolumntype{C}[1]{>{\centering\let\newline\\arraybackslash\hspace{0pt}}m{#1}}
\newcolumntype{R}[1]{>{\raggedleft\let\newline\\arraybackslash\hspace{0pt}}m{#1}}

\begin{document}

\begin{titlepage}
\title{Racial Disparities in Debt Collection}

\author{Jessica LaVoice\thanks{Department of Economics, Bowdoin College, 255 Maine St, Brunswick, ME 04011. Email: jlavoice@bowdoin.edu. (Corresponding Author)} \hspace{3mm} Domonkos F. Vamossy\thanks{Department of Economics, University of Pittsburgh, 4200 Fifth Ave, Pittsburgh, PA 15260. Email: d.vamossy@pitt.edu.}}
\date{June 2023}
\maketitle

\noindent
%\large \textbf{This is a preliminary draft. Please do not cite or circulate without permission from the author.}

\begin{abstract}
	\singlespacing
\noindent 

This paper shows that black and Hispanic borrowers are 39\% more likely to experience a debt collection judgment than white borrowers, even after controlling for credit scores and other relevant credit attributes. The racial gap in judgments is more pronounced in areas with a high density of payday lenders, a high share of income-less households, and low levels of tertiary education. State-level measures of racial discrimination cannot explain the judgment gap, nor can neighborhood-level differences in the previous share of contested judgments or cases with attorney representation. A back-of-the-envelope calculation suggests that closing the racial wealth gap could significantly reduce the racial disparity in debt collection judgments. 

%\vspace{0in}\\
%\vspace{0in}\\

%D11	Consumer Economics: Theory
%D12	Consumer Economics: Empirical Analysis
%D14 Household Saving; Personal Finance
%D18	Consumer Protection
%D63 Equity, Justice, Inequality, and Other Normative Criteria and Measurement
%D82 Asymmetric and Private Information; Mechanism Design
%G21 Banks; Depository Institutions; Micro Finance Institutions; Mortgages
%G28	Government Policy and Regulation
%J15 Economics of Minorities, Races, Indigenous Peoples, and Immigrants • Non-labor Discrimination
%L84 Personal, Professional, and Business Services

%O33 Technological Change: Choices and Consequences; Diffusion Processes

 \bigskip
\end{abstract}

\setcounter{page}{0}
\thispagestyle{empty}
\end{titlepage}
\pagebreak \newpage

\doublespacing

\input{Sections/introduction}

\input{Sections/background}

\input{Sections/data}

\input{Sections/methods}

\input{Sections/results}

\input{Sections/conclusion}

\pagebreak

\singlespacing
\nocite{*}
\bibliographystyle{plainnat}
\bibliography{bib}

\clearpage

\input{Sections/figures}

\input{Sections/tables}

\clearpage

\doublespacing

\input{Sections/appendix}

\end{document}

%% file: Sections/introduction.tex
\section{Introduction} \label{sec:introduction}

Debt collection court cases have significant financial implications for debtors. For instance, a guilty verdict, referred to as a judgment, can result in wage garnishments and seizures of bank accounts. These consequences, particularly when compounded with interest and legal fees, hinder debtors' ability to achieve financial stability and limit their ability to accumulate wealth. Moreover, judgments disproportionately occur in predominantly Black neighborhoods, even after controlling for income differences across neighborhoods (Keil and Waldman, 2019).\footnote{More specifically, the risk of judgment is twice as high in majority-Black neighborhoods than in non-majority Black neighborhoods with similar income levels (Keil and Waldman, 2019).} 

In this paper, we document a racial disparity in the likelihood of receiving a debt collection judgment after accounting for potential confounding factors including credit scores, debt levels, and other debt characteristics.  To do so, we create a nationally representative individual-level panel dataset spanning from 2013Q1 to 2017Q2 containing both individuals demographic information as well as a comprehensive set of their credit report characteristics.  Credit information comes from Experian credit reports and includes indicators for if an individual has any outstanding judgments. While credit reports do not contain information about an individual's race, following the methodology in \cite{butler2022racial}, we merge credit report data with information from the Home Mortgage Disclosure Act (HMDA) database to uncover borrowers' race and ethnicity. We find that the judgment rate is 39\% higher for black and Hispanic borrowers, controlling for credit scores, debt levels, and other debt characteristics.  This disparity is not correlated with measures of racial bias, but is instead driven by locations with a high number of payday lenders, a high share of income-less households, and low levels of tertiary education. Conditional on a judgment being issued, we find no statistical difference in the likelihood a judgment is petitioned or satisfied across minority and non-minority borrowers, suggesting limited evidence that collectors are focusing their collection efforts based on the probability a debt will be petitioned or that the account will be brought to current.    

We further examine potential mechanisms through which this disparity could permeate the debt collection system. One potential explanation for the gap in judgments is differences in neighborhood-level characteristics that impact the probability of success and profitability of collecting on a debt. When dealing with unpaid balances on unsecured debt, creditors have two main options if negotiating with the borrower has proved unsuccessful: they can write off the debt or take the debt to collections. Several factors drive this decision, including the likelihood the debtor will ultimately repay the debt and the cost of collecting on the debt. While creditors are legally prohibited from using race as a factor when making decisions related to credit, they may use neighborhood characteristics to gather information about a debtor’s ability to repay debt or to estimate the potential cost of collecting the debt. For example, it could be the case that defendants from black neighborhoods are less likely to hire an attorney or to contest the debt in court, making it less costly for debt collectors to obtain judgments in black neighborhoods. While creditors do not know this information a priori for any given individual, they could estimate the likelihood of such situations based on historical outcomes in a given zip code. 

To analyze the extent to which such mechanisms are driving the racial disparity, we create a zip code-level panel dataset spanning from 2004 to 2013 that links the number of debt collection judgments in each zip code in Missouri with data from Experian credit reports and the American Community Survey. Controlling for a comprehensive set of financial variables derived from the credit report data to account for all aspects of neighborhoods' financial liabilities, we find that debt collection court cases are 56\% more common in black neighborhoods compared to non-black neighborhoods. Using this data, we find no statistical difference in the share of contested judgments across black and non-black communities, and controlling for attorney representation has a limited effect on the judgment gap. 

Another potential mechanism that could explained the racial gap in debt collection judgments results from the broader disadvantages experienced by black communities. According to estimates provided by the United States Census Bureau in 2016, the typical black household has a net worth of \$12,920, while that of a typical white household is \$114,700 - this is a \$101,780 difference in wealth that could have important implications for a household’s ability to mitigate negative financial shocks. About \$35,000 of this wealth gap is not driven by home equity and would not be captured in our current analysis. By translating this wealth gap into differences in annual income and using our estimates of the relationship between income and judgments from our matched credit bureau data, we calculate that a wealth gap of this size would explain the judgment gap.\footnote{Our estimated racial judgment gap suggests black and Hispanic individuals are 0.0496 percentage points more likely to have a new judgment on their credit report, holding credit score, income, and other credit characteristics constant. We computed the difference in annual savings needed over a 40-year horizon to generate a wealth gap of \$35,000. We found that an annual difference of \$2,910 is sufficient to generate the wealth gap in net present value. For the interest rate, we applied the stock market’s historical return, which between 1957 through 2018 is roughly 8\%. Consistent with estimates from the U.S. Bureau of Economic Analysis, we assume an 8\% personal savings rate. Under these assumptions, this wealth gap translates into an annual income difference of \$36,375. Using our estimated regression equation, we conclude that increasing the imputed income of minority individuals by this amount would eliminate the judgment gap, while increasing the HMDA income by this amount would explain 77\% of the judgment gap.}

This paper contributes to an extensive literature that documents the role that race plays in the legal, financial, and criminal justice system. For example, black individuals are more likely to be searched for contraband (\cite{antonovicsknight2009}), to have biased bail hearings (\cite{Arnoldetal2018}), and to be charged with a serious offense (\cite{MaritStarr2014}). Black and Hispanic borrowers are less likely to get approved for an auto loan, even after controlling for creditworthiness (\cite{butler2022racial}), and  African American zip codes are less likely to receive financial restitution from complaints submitted to the Consumer Financial Protection Bureau (\cite{haendler2021financial}).  Discrimination in the labor and housing-related market has also been well documented (e.g., \cite{bartlett2019consumer}, \cite{ritter2011racial}, \cite{bertrand2004emily}, and \cite{turner2002discrimination}). Such disparities have contributed to racial differences in wealth, and a growing literature explores how the racial wealth gap was generated and how it has persisted over time (e.g., \cite{mckernan2014racial}; \cite{akbar2019racial}; \cite{derenoncourt2022wealth}). While much attention has been given to racial disparities in general, this is the first economic analysis to empirically document racial disparities in debt collection judgments.  

Aside from the literature documenting racial disparities and discrimination, this paper also contributes to a growing literature about the debt collection industry.\footnote{See \cite{hunt2007} for an overview of the debt collection industry and details about its institutional structure and regulatory environment.}  We know that consumers whom creditors and debt collectors sue are drawn predominantly from lower-income areas (\cite{hynes2008}).  Other more recent studies have investigated the role of information technology in the collection of consumer debts (\cite{drozd}), documented the link between debt collection regulations and the supply of consumer credit (\cite{Fedasseyeu2015}), and determined if consumers are made better or worse off by settling their debt outside of court (\cite{Chengetal2019}).  Interpreted broadly, the literature has documented the impacts of the debt collection process on consumer outcomes; however, racial disparities in debt collection have not been rigorously explored.    

The rest of our paper is outlined as follows:  Section 2 discusses the typical debt collection litigation process in the United States, as well as the laws regulating access to credit and debt collection procedures; Section 3 describes our data; Section 4 discusses our empirical approach; Section 5 documents the racial gap in debt collection judgments and explores potential mechanisms that may be driving the disparity; Section 6 concludes.

%% file: Sections/background.tex
\section{Background Information}\label{sec:information}

The debt collection industry in the United States is large and growing.  According to a 2018 annual report by the Consumer Financial Protection Bureau, debt collection is a \$10.9 billion industry that employs nearly 120,000 people across approximately 8,000 collection agencies in the United States.  In 2010 alone, U.S. businesses placed \$150 billion in debt with collection agencies.  When the debt is unsecured, the owner of the debt can either negotiate with the debtor to bring their debt to current, write off the debt, or file a debt collection lawsuit.  This section will summarize the key institutional details surrounding debt collection lawsuits and the laws regulating the debt collection industry.  

\subsection{Debt Collection Litigation Process}

Debt collection litigation typically begins when a creditor files a ``Summons and Complaint'' in a state civil court.\footnote{These courts have many different names, including municipal court, superior court, justice court, county court, etc.}  This document names the parties involved and states the amount owed (including interest and, in some cases, attorney fees and court costs).  The summons is served to the defendant to notify them of the lawsuit.  It also provides the defendant with additional information, including the deadline for which the debtor must file a formal response to the court.  If the debtor does not meet this deadline, the creditor will usually ask the court to enter a default judgment, at which point the defendant is obligated to abide by the court's ruling and is subject to the punishments requested by the court.  

For most routine debt collection lawsuits, if the debtor files a formal response to the lawsuit, a trial date will be requested and set by the court.  In some courts, settlement conferences are held to provide both parties with the opportunity to settle the case before the trial.  Once the creditor obtains a judgment, the creditor might request a ``debtor's examination,'' which would require the debtor to appear in court and answer questions about their finances.  This process informs the creditor how it can collect the judgment.  The most common methods for enforcing the judgments are to garnish wages or bank accounts.\footnote{Courts can also seize and sell the debtor's personal property, though this is relatively uncommon.}  If a dispute is settled before trial, the creditor gives up the ability to collect on the debt by garnishing the debtor's bank accounts or wages, and therefore creditors often require a one-time lump sum payment to drop the suit. 

\subsection{Laws Regulating Debt Collection}

Debtors are granted some protections throughout the debt collection process. The 1977 Fair Debt Collection Practices Act (FDCPA) is the primary federal law governing debt collection practices. The statute's stated purposes are as follows: to eliminate the abusive practices used to collect consumer debts, such as calling the debtor at all hours of the night and showing up to their place of employment, to promote fair debt collection, and to provide consumers with an avenue for disputing and obtaining validation of debt information to ensure the information's accuracy.  

The Consumer Credit Protection Act (CCPA) of 1968 restricts the amount of earnings that creditors can garnish from defendants' weekly disposable income to 25\% or the amount by which disposable earnings are greater than 30 times the minimum wage. State laws can increase the share of wages protected from debt collection garnishments. For example, a creditor in Missouri can only garnish 10\% of after-tax wages if the debtor is the head of their household. However, the burden to assert these protections is typically on the debtor, and take-up is relatively low. No federal law limits the amount of savings that can be seized from a debtor's bank accounts.    

While not directly related to debt collection, other protections have been put in place to protect consumers in the credit market. For example, the Equal Credit Opportunity Act (ECOA), enacted in 1974, makes it illegal for creditors to discriminate against any applicant based on race, color, religion, national origin, sex, marital status, age, or participation in a public assistance program.\footnote{This law is enforced by the Federal Trade Commission (FTC).} The law applies to everyone who regularly participates in a credit decision, including banks, retail and department stores, bankcard companies, finance companies, and credit unions. The ECOA applies to the decision to grant credit and set credit terms. 

Furthermore, the Fair Credit Reporting Act (FCRA) of 1970 promotes the accuracy, fairness, and privacy of consumer information contained in the files of consumer reporting agencies. The law protected consumers from including inaccurate information in their credit reports. The Credit Card Accountability Responsibility and Disclosure (CARD) Act of 2009 established fair and transparent credit card practices. Key provisions include giving consumers enough time to pay their bills, prohibiting retroactive rate increases, making it easier to pay down debt, eliminating ``fee harvester cards'', and eliminating excessive marketing to young people. Despite these protections, abusive debt collection practices still exist, and, as we will show, black and Hispanic borrowers are disproportionately impacted by debt collection judgments.

%% file: Sections/data.tex
\section{Data} \label{sec:data}

This paper uses two different datasets to explore racial differences in debt collection judgments. The first is an individual-level panel dataset that combines financial information from individuals' Experian credit reports with demographic data from the Home Mortgage Disclosure Act (HMDA) database. This dataset is used to document the racial disparities in debt collections judgments and to identify where the disparity is the largest.  The second is a zipcode-level panel dataset containing debt collection court cases filed in Missouri between 2004 and 2013, along with other neighborhood characteristics.  This dataset is used to explore the extent to which neighborhood debt collection characteristics are driving the racial gap in judgments.

\subsection{Credit Report/HMDA Individual Level Panel}

To construct the Credit Report/HMDA matched panel, we start with a panel data set of credit bureau records. The Credit Report data is an anonymous, nationally representative, individual-level panel dataset from 2004Q1 to 2017Q2 with quarterly observations. This data contains a random draw of over half a million individuals with an Experian credit report and consists of over 200 variables that allow us to track all aspects of individuals' financial liabilities, detailed delinquencies, various types of debt with the number of accounts and balances, credit score, and an imputed income estimate. Importantly, this data also indicates the number of judgments outstanding against a given individual in a given quarter from 2013Q1 to 2017Q2.  Also included is where a judgment has been petitioned or satisfied. 

While this data does include an individuals age and zipcode, the Equal Credit Opportunity Act (ECOA) generally prohibits a creditor from inquiring about the race, color, religion, national origin, or sex of an applicant. As such, many data sets, including the Experian credit report data, lack information about an applicant's race. To obtain this information, we match the Experian credit report data with demographic information from the Home Mortgage Disclosure Act (HMDA) database.  The Home Mortgage Disclosure Act mandates that almost all mortgage lenders report comprehensive information on the applications they receive and whether they approve the loan.\footnote{Exceptions are granted only for very small or strictly rural lenders (\cite{butler2022racial}). Approximately 95\% of all first-lien mortgages are reported to the HMDA database (\cite{avery2017profile}), with the coverage rate likely being higher for properties in MSAs.} Mortgage lenders submit applicants' racial and ethnic background, personal attributes, and loan application details, such as the requested loan size, income, loan purpose (purchase, refinancing, improvement), co-applicants, loan priority (first or second lien), and the census tract location of the property, to the HMDA database. If a loan is granted, any loan sale is reported, along with an indicator for sale to any quasi-government entity. Our HMDA data ranges from 2007 to 2017.

Both the credit bureau and HMDA data are anonymized, with no unique identifier linking the two datasets; however, the detailed information on originated mortgages in both datasets allows mortgages to be matched based on their features (\cite{butler2022racial}). Following \cite{butler2022racial}, we first apply a set of filters to the mortgages from our credit bureau data. We require the mortgage to be the borrower's only first lien mortgage at the time to ensure that the borrower's location in the credit bureau data will match the property location in the HMDA data. We also require that the person lives in an MSA directly following the loan origination and that they were the only applicant on the loan. We also follow the approach of \cite{butler2022racial} for processing the HMDA data. We concentrate on home purchase and refinancing loans (excluding home improvement loans due to their less defined nature in both datasets). We mandate mortgages to be on owner-occupied properties to ensure that the property location aligns with the borrower's location in the credit bureau data. We also stipulate that the mortgage must be a first lien and the property must be situated within an MSA (where HMDA data is most comprehensive). Lastly, we require that the mortgage have only one applicant/borrower, so the demographic data directly applies to the matched individual in the credit bureau data.  

After applying the filters to both the HMDA and credit bureau data, the target population for the matched sample comprises borrowers obtaining a home purchase or refinance loan individually (no co-applicant), for their primary residence located within an MSA, between 2007 and 2017. We matched the credit bureau data to HMDA records following the procedure outlined in \cite{butler2022racial}, with slight modifications. Because our Experian credit report data only reports an individuals zip code, we first matched our zip code level credit bureau data to Census 5-Digit Zip Code Tabulation Areas (ZCTA5s) using the Missouri Census Data Center's Census Tract - ZIP/ZCTA crosswalk files.\footnote{USPS zip codes are not areal features used by the Census but a collection of mail delivery routes that identify the individual post office or metropolitan area delivery station associated with mailing addresses. ZIP Code Tabulation Areas (ZCTAs) are generalized areal representations of United States Postal Service (USPS) ZIP Code service areas. HMDA data reports 2010 census tracts starting in 2013. For loans originated from 2003 to 2012, HMDA data uses the 2000 census tracts. We successfully matched ZCTAs for 96.3\% of Census Tracts in the HMDA data. Crosswalk URL:  \url{https://mcdc.missouri.edu/applications/geocorr2014.html}.} This process increased our HMDA sample from 28.4M to 50.7M observations as it assigns each individual to all potential ZCTA5s.

We matched mortgages in the credit bureau data to HMDA data based on six characteristics: origination year, ZCTA5, loan amount, loan purpose (purchase or refinancing), mortgage type (conventional, FHA, or VA), and if/to which quasi-government entity the loan is sold. We implement an exact matching algorithm, as opposed to using nearest matches or propensity scores, to ensure the highest accuracy for our matched dataset.\footnote{In our analysis of home purchase loans, we focus on borrowers without an outstanding first-lien mortgage balance from the previous two quarters. To mitigate the impact of integer-based origination balances in HMDA data, we consider credit bureau data origination balances within a specific range. By allowing neighboring integer values, we prevent inaccuracies caused by rounding discrepancies. All other matching variables are exact matches.}  Replacing ZCTA5 with census tracts, 90.4\% of originated mortgages in the HMDA data are unique based on these characteristics. Before expanding census tracts to ZCTA5, we removed duplicate entries. After expanding the census tracts to ZCTA5 and prior to merging with credit bureau data, we filtered out observations with identical race, gender, ethnicity, and mortgage origination details. These entries appear because we expanded our census tract data to the ZIP code level. This filter and the removal of duplicates based on our matching variables result in a sample of 39.7M observations. 

\cite{butler2022racial} identifies two potential sources of error when matching HMDA with credit bureau data. First, a data error in one of the matching variables might lead to a mismatch. Second, HMDA-reporting and non-reporting lenders may originate identical, otherwise unique mortgages. They argue that such mismatches should not introduce bias in estimates beyond pure noise. Our matching process, which expands HMDA data from census tract to ZIP code level, introduces two additional error sources. First, copies created by assigning an individual from a given census tract to multiple zipcodes may find a matching observation in the incorrect zip code. This would be a severe problem if the correct loan was not observed in our HMDA data. However, the HMDA data covers 90\% of unique mortgages at the census tract level, and hence we likely observe the correct loan as well. Second, two loans from different census tracts with identical matching variables could be copied to the same ZIP code. This issue affects 34\% of HMDA observations. This is slightly more likely in census tracts with diverse populations. We demonstrate this in Figure \ref{fig:deletion_by_majority_share}, which shows the probability of having duplicated observations along our match variables after expanding the census tract level data to ZCTA5. To address this problem, we remove observations with conflicting protected attribute information at the match level.

We report the match rate as well as summary statistics on the match in Table \ref{table:1}. Panel A displays the match rate for home purchase mortgages, refinance loans, and both types of loans combined. We find a corresponding HMDA mortgage for 45.6\% of the mortgages in the credit bureau data with a slightly higher match rate for refinance loans as opposed to home purchase loans. Given that HMDA covers 95\% of mortgages, 90\% are unique at the census tract level, 96\% of census tracts are merged to ZCTA5, and 66\% of observations are unique based on our matching variables, the best achievable match rate is approximately 54.2\%. In other words, our algorithm successfully found matches in roughly 83.9\% (0.455/0.542) of the potential cases. 

The summary statistics in panels B and C of Table \ref{table:1} assess whether our matched sample is representative of the original population for home purchase loans and refinance loans respectively. Panel B reveals that the successfully matched home purchase mortgage sample is largely representative of the initial population of credit bureau mortgages. Similarly, Panel C demonstrates that the matched sample of refinance loans accurately represents the starting sample from the credit bureau data.

Since our study focuses on the impact of race on debt collection judgments, it is essential to determine if minorities are underrepresented in the data or if a specific type of minority borrower, such as high or low income, is underrepresented. Following \cite{butler2022racial}, we investigate whether race plays a role in the likelihood of matching originated mortgages from the HMDA database to our sample of credit bureau records. The regressions in Table \ref{table:2} assess the probability of successfully matching a loan based on borrower race. The coefficients on Black and Hispanic, and the interaction terms Black X log(Income) and Hispanic X log(Income) in column (1) are insignificant. 
While there is a slight indication of selection bias through the combination of race and income for Hispanic applicants seeking refinance mortgages, we can control for variables from both databases, which mitigates concerns about selection bias.

Table \ref{table:3} provides summary statistics for our individual-level panel data starting in 2013Q1 and ending in 2017Q2.\footnote{While a larger sample of data is used to match credit report and HMDA data, we limit our sample for our regression analysis as judgment data isn't reliably available in every quarter of the credit report data.} Similar to patterns identified in \cite{butler2022racial}, Columns 1 and 2 indicate that individuals within the matched data set generally exhibit higher credit scores, are relatively younger, and have a higher likelihood of holding a mortgage compared to the average U.S. resident who has a credit history. These patterns are anticipated, as individuals must either obtain a new mortgage or refinance an existing one between 2007 and 2017 to qualify for inclusion in the matched panel. Columns 3–5 demonstrate that the White borrowers in the matched panel have higher credit scores and incomes in comparison to minority (non-Asian) borrowers. 

We supplement our matched Credit Bureau/HMDA panel with data on racial biases in U.S. states based on the Google Search Volume for racial slurs, adopting the approach of \cite{stephens2014cost}. We also use Census ZIP Code Business Patterns (ZCBP) data to access the number of banks and payday lenders in each zip code. Following \cite{Bhutta2014}, we use the following two North American Industrial Classification System (NAICS) codes to capture payday lending establishments:  non-depository consumer lending (establishments primarily engaged in making unsecured cash loans to consumers) and other activities related to credit intermediation (establishments primarily engaged in facilitating credit intermediation, including check cashing services and money order issuance services).\footnote{\cite{Barthetal2016} discuss how this proxy could likely overstate the number of payday lenders. We adjust these measures to correct for states that prohibit payday lending to help reduce this bias.}  For each zip code, we use ArcGIS to create a weighted average (based on land area) of the number of banks and payday lenders within a five-mile radius of each zip code's centroid. Lastly, we add data from the 2009-2013 American Community Survey to control for population, education, and income at the zip code level.

\subsection{Missouri Neighborhood-Level Panel}

Our second dataset is a panel dataset containing zip codes in Missouri.\footnote{The Missouri court data was generously provided by Kiel and Waldman (2015). They acquired individual court case data from the state court administration. Their white paper focuses on three jurisdictions: Cook County, Illinois (composed of Chicago and surrounding suburbs), St. Louis City and St. Louis County, Missouri, and Essex County, New Jersey (composed of Newark and suburbs). The court data included basic case information such as the plaintiff, the defendant, and the defendant’s address. The defendant’s race is not reported. See Appendix section \ref{appendixa1} for additional data details.}  Aside from the fact that Missouri has a centralized database of cases tried in different circuit courts, we focus on Missouri due to it's representative nature in terms of collection (\cite{Ratcliffeetal2014}; \cite{Chengetal2019}). More specifically, Missouri is representative regarding the percentage of delinquent consumers and the average amount of debt in collections (\cite{Ratcliffeetal2014}). Missouri is also not particularly exceptional regarding its law surrounding collections, its share of black residents, or its level of inequality (\cite{Chengetal2019}).\footnote{We have additional debt collection judgment data from New Jersey and Cook County, Illinois (composed of Chicago and surrounding suburbs); however, it is less detailed than the Missouri data. While this data includes the number of cases filed in a five-year window from 2008-2012, it is not broken down by type of judgment (default, consent, contested), defendants' names, or representing attorney. As such, all of our main specifications use data from Missouri due to its representative nature, high level of detail, and to keep our sample consistent across specifications. However, analogous analysis is provided in the appendix when comparable information is available.} 

This dataset includes debt collection court cases filed in Missouri between 2004 and 2013, along with other neighborhood characteristics.  For each zip code in our sample, we know the number of debt collection lawsuits filed and the number of judgments arising from these lawsuits. We also see the number of cases that resulted in a default judgment (meaning the debtor did not show up to court), a consent judgment (meaning the debtor showed up to court and admitted to owing the debt), and the number of cases that were contested (meaning that some aspect of the debt was disputed). We know if the defendant was represented by an attorney and the plaintiff type (categorized into the following groups: auto, debt buyer, high-cost lender, major bank, medical, utility, and miscellaneous).

The Missouri court data is merged with Experian credit report data aggregated to the zip code-level to control median credit score, the average number of delinquent accounts, and other neighborhood-level credit measures for each zip code in our sample. Zip code tabulation data from the IRS from 2004 - 2018 is used to control for the average income and the fraction of all filers with income under \$25,000, \$25,000-50,000, \$50,000-75,000, \$75,000-100,000 and over \$100,000.  Also included is zip code tabulation data from the 2009-2013 American Community Survey to control for population, racial composition, the unemployment rate, and other socioeconomic variables of interest at the zip code-level. 

For our neighborhood-level analysis, our outcome variable of interest is the number of judgments per 100 people. Neighborhood racial composition is our primary independent variable of interest. More specifically, we classify a zip code as a majority-black if more than 50\% of its residents are black.\footnote{Section A.2.6 in the appendix shows that our results are robust to other measures of neighborhood racial composition, including using the share of the population that is black.} Throughout our analysis, we control for other neighborhood-level characteristics that may be driving the racial gap in debt collection judgments. For example, we document the share of cases in which an attorney was present as well as the share of cases in which the debt was contested.  

We limit our sample to only observations with common support over observables (\cite{crump2009dealing}).\footnote{Details regarding the creation of this sample are discussed in the section \ref{appendixa1} of the appendix. See Table \ref{tab:sample_restrictions} in the appendix for sample size restrictions based on data availablity and restricting the sample to the common support.} Table \ref{tab:summary_statistics} summarizes neighborhood-level data from over 800 Missouri zip codes from 2004-2013. This data shows that majority-black neighborhoods have higher judgment rates (2.7 per 100 people) than majority non-black neighborhoods (1.3 per 100 people). The data also suggests that majority-black neighborhoods have lower incomes and credit scores, higher past-due debt balances, more default judgments, fewer attorney-represented defendants, more payday lenders, lower house values, lower homeownership rates, higher unemployment rates, and higher divorce rates. Most differences are significant at the 1\% level.

%% file: Sections/methods.tex
\section{Empirical Methods} \label{sec:methods}

Focusing first on the individual-level matched credit bureau data, we investigate racial disparities in debt collection by regressing judgment outstanding on minority, an indicator for the person being Black or Hispanic. Our empirical specification is given by the following equation:

\begin{equation}
    y_{izt}=\alpha + \beta M_{iz} + \theta X_{izt} + \gamma_z + \lambda_{st} + \epsilon_{izt}
\end{equation}

\noindent
where $y_{izt}$ is a binary variable indicating if individual i in zipcode z has a new judgment on their credit score in quarter t, $M_{iz}$ is an indicator variable equal to one if individual i in zipcode z is Black or Hispanic, and $X_{izt}$ is a vector of other controls for individual i in zipcode z in quarter t. Such controls include credit score, measures of debt balances by type of debt (credit card, mortgage loans, student loans, etc.), debt composition (type of debt as a share of total debt balances), delinquent balances by the length of delinquency (30 days, 60 days, 90 days), and bankruptcy/collection flags and indicators for the timing relative to the borrower's credit bureau/HMDA match, and imputed income/HMDA income wherever applicable. In addition, our main specification includes zipcode fixed effects to control for time-invariant differences across zipcodes and state-by-quarter fixed effects to control for any time-varying changes that impact all individuals within a state. 

We run a similar specification using our neighborhood level dataset from Missouri. Our outcome of interest becomes the number of judgments per every 100 people in zip code i in county c in year t.  Our control variable of interest is an indicator variable for if the neighborhood is a majority black neighborhood (defined as a neighborhood where the black population is greater than 50\%). We control for neighborhood-level income and credit score measures, measures of debt balances by type of debt, debt composition, delinquent balances by the length of delinquency, and bankruptcy/collection flags. We also control for other neighborhood level observable characteristics such as the Gini index, unemployment rate, median housing values, education levels, and divorce rate. All regressions include county and year fixed effects. To better understand the mechanisms driving the racial disparity in debt collection judgments, we include a number of additional control variables at the neighborhood level. Such control variables include the previous share of cases in which an attorney was present as well as the previous share of cases in which the debt was contested. All regressions are weighted by population in our neighborhood-level analysis. Standard errors are clustered at the county-year level and we report the p-values from a wild cluster bootstrap for inference due to the small number of clusters (Cameron and Miller, 2015).

%% file: Sections/results.tex
\section{Results} \label{sec:result}

In section \ref{mainresults} we document a racial disparity in debt collection judgments using our Credit Report/HMDA individual-level panel. We conclude that debt collection judgments are 39\% more common among black and Hispanic individuals. We then present our neighborhood-level results to better understand potential mechanisms that could be driving this disparity.

\subsection{Racial Disparity in Judgements}\label{mainresults}

We begin by estimating equation (1) to determine the extent to which there is a racial disparity in debt collection judgments.  Table \ref{table:4a} presents these regression results.  Column (2) of Table \ref{table:4a} shows that minority individuals experience a 0.1156 percentage point increase in the probability of having a new judgment on their credit report after controlling for state-by-quarter fixed effects, zip code fixed effects, and demographic information. After adding credit report controls to the previous specification, we find that minority individuals are 0.0496 percentage points more likely to accumulate judgments (Column (3)); this is a 39\% increase from the judgment rate among white borrowers.  These results highlight both a substantial gap that cannot be solely explained by differences in credit characteristics as well as the importance of including measures of credit quality in our analysis. This unexplained difference in judgments is roughly the same size as we would see from a 21-point (24\% of a standard deviation) reduction in credit score. In column (4) we include imputed income derived from the credit bureau data and income from HMDA.  The coefficient on minority only decreases by 10\%.  

The information presented in columns (5) and (6) of Table \ref{table:4a} indicate that minorities tend to accrue new judgments more frequently as subprime borrowers in absolute terms, but not proportionately. The coefficient pertaining to minorities who are prime borrowers is borderline insignificant, suggesting a potential judgment rate that is twice that of non-minority borrowers. This particular observation carries significant weight. The decision-making process for subprime borrower approval typically involves a higher degree of discretion by loan officers. This can lead to a reduction in the marginal cost associated with decisions that have potential discriminatory implications.

In Panels A and B of Table \ref{table:4bc} we change our dependent variable to whether or not a judgment was reported as being petitioned or satisfied, respectively. We find no statistically significant racial disparities in the number of judgments petitioned or satisfied once conditioning on having an outstanding judgment in the previous quarter. These findings suggest that debt collectors have no incentive to target specific segments of the population based on their likelihoods to petition or satisfy their judgment.

We next explore where racial disparities in debt collection judgments are most pronounced. Initially, we investigate if these discrepancies are more significant in states with high racial bias, similar studies by \cite{charles2008prejudice} and \cite{butler2022racial}. We measure racial biases by adopting the methodology of \cite{stephens2014cost}, which uses the volume of Google Search for racial slurs. A crucial advantage of leveraging Google searches is that they reveal attitudes individuals typically hide in survey responses. In column (2) of Table \ref{table:5}, we find that the interaction between minority status and a binary variable for high racial bias is not statistically significant. This implies that the racial disparities observed in debt collection judgments cannot be attributed to state-level taste-based discrimination. 

Our next cross-sectional test is based on the prevalence of non-bank lending in the ZIP code where the applicant lives. In column (3) of Table \ref{table:5}, we find that the effect of race on judgment rates is significantly different for applicants in zip codes in the top quartile of bank lending share (i.e., areas with less reliance on non-bank lending, like payday lenders) compared to the remaining ZIP codes. This finding suggests that racial disparities in judgment rates are more prevalent in areas with a higher fraction of payday lenders. It is important to note that payday lenders are often located in areas with high financial hardship, serving communities where traditional banking services may be less accessible, and people may have limited financial options. This tends to disproportionately affect minority communities, potentially contributing to the disparities observed in judgment rates. High judgment rates and the prevalence of payday lenders are likely symptoms of underlying socioeconomic disparities, rather than the presence of payday lenders causing higher judgment rates among minorities. 

Viewing the issue from a different perspective, Column (4) looks at ZIP codes that fall within the top quartile of households without income and discovers that the disparity originates from individuals residing in these areas. Conversely, Column (5) examines individuals living in ZIP codes in the top quartile of tertiary education rates and finds no observable disparities within these regions. 

\subsection{Neighborhood Level Mechanisms}\label{mechanisms}

Why are minorities more likely to experience a debt collection judgment than white individuals, despite having similar credit characteristics? It could be that creditors are using neighborhood-level information to help determine the profitability of their debt collection efforts\footnote{Creditors can legally factor proxy variables into their decision-making if there is a legitimate business necessity, such as scoring credit risk, even if these tactics cause disparate outcomes across space (\cite{bartlett2021consumer}).} For example, they may focus their collection efforts in neighborhoods where defendants are less likely to hire an attorney or to contest the debt in court. This section uses neighborhood-level data from Missouri to explore if and to what extent neighborhood-level characteristics can explain the racial disparity in debt collection judgments. We begin by replicating the findings from the individual analysis, which used nationally representative judgment data from Experian, with judgment data from Missouri aggregated to the zip code level. We then explore neighborhood-level characteristics that could drive the relationship between neighborhood racial composition and debt collection judgments. 

We start by documenting the racial disparity in judgments across black and non-black neighborhoods.  Figure \ref{fig:demographics} uses Missouri neighborhood data, grouping zip codes into 100 bins by the share of black residents. The figure plots the average black population share against the average judgment rate per bin, with bubble size denoting the number of zip codes per bin. The regression line, weighted by observation count, shows a positive correlation between judgment rate and black population share.

Figure \ref{fig:cs_vs_income} explores the correlation between racial composition, median income, credit scores, and debt judgments in neighborhoods. Panel (a) indicates a negative correlation between judgment rate and median income, even when comparing neighborhoods of similar income levels, with higher judgment rates seen in majority-black neighborhoods. Panel (b) shows a negative correlation between judgment rate and median credit scores, with higher judgment rates in majority-black neighborhoods despite similar credit scores. This suggests differences in income and credit scores may not be the primary mechanism driving the racial disparity in debt collection cases.  

We next regress the number of judgments per 100 people on an indicator for if a neighborhood is majority black, as well as average ZIP code level income and the fraction of IRS filings under \$25,000, between \$25,000-50,000, \$50,000-75,000, \$75,000-100,000, and over \$100,000, credit score quintiles and median credit score to the baseline specification, total delinquent debt balances, unemployment rate, median house value, the fraction of the population with a college education, the divorce rate, and population density. Results are presented in Table \ref{tab:debt_composition}. Our preferred specification is given in column (5) and includes all our controls for individuals debt portfolios, although the coefficient is positive and statistically significant across each specification. Column (5) suggests there are approximately 0.70 more judgments per 100 people in majority black zip codes compared to majority non-black neighborhoods; this translates in a 56\% higher judgment rate.

One potential difference in the cost of collecting debt across black and non-black neighborhoods is the likelihood that an attorney represents a given debtor. While this is not known a priori, creditors could theoretically estimate these likelihoods when deciding which delinquent accounts to bring to court. To explore how differences in attorney representation drive our results, we explore the disparity in attorney representation and investigate whether this impacts our main findings. These results are presented in Columns (1) and (2) of Table \ref{tab:judgment_type}.  Each specification in this table includes the income, credit score, and baseline controls discussed above, as well as county and year fixed effects. The outcome variable in Column (1) is the share of debt collection court cases where an attorney represented the defendant; this result shows that defendants in majority-black neighborhoods are less likely to have an attorney represent them in a debt collection court case. However, as seen in Column (2), where our dependent variable is once again judgments per 100 people, controlling for the share of cases in which an attorney represents defendants cannot explain the racial disparity in debt collection cases.\footnote{This does not imply that attorney representation is not meaningful or important in debt collection court cases. Debt collection laws often place the burden on asserting various legal protections, including the share of the debtor's wages that can be garnished as the result of a judgment on the debtor. Thus, attorney representation is important in protecting debtors' rights throughout the debt collection process, even if such cases ultimately end in judgments.} 

It could be the case that debt collectors target their collection efforts in areas where defendants are less likely to show up to court, resulting in a default judgment, or in areas where defendants do not tend to contest the debt in court. In other words, debt collectors might avoid collecting in areas where defendants tend to argue some aspect of the debt owed, which could result in the plaintiff exerting more effort or spending more money to collect the debt.  

To explore how differences in the share of contested versus uncontested cases could drive our result, we document the impact of neighborhood racial composition on the share of different types of judgments. Our outcome variable in Column (3) Table \ref{tab:judgment_type} is the share of cases in which the defendant admitted to owing the debt, and our outcome variable in Column (4) is the share of cases that were contested. We see no racial differences along these dimensions. Columns (6), (7), and (8) explore the share of cases that resulted in default judgments, were dismissed, or were settled, respectively. We see that a larger share of cases in black neighborhoods resulted in default judgments or was dismissed, while a smaller share was settled before court.

Given that the rank order preference of case outcomes is not apparent, we explore how these differences could impact the judgment gap by including the lagged share of case outcomes into our main specification, where our outcome variable is judgments per 100 people.  Including lagged case outcomes does not mitigate the judgment gap across black and non-black neighborhoods. These results suggest that previous differences in case outcomes across neighborhoods and plaintiffs' potential strategic decisions to exploit this information for cost savings measures do not explain why judgments are more common in black neighborhoods.\footnote{Section \ref{appendixa2} in the appendix contains a multitude of additional analysis and robustness checks. For example, we show the racial gap in debt collection judgments exists for every plaintiff type (albeit at different magnitudes) except medical lenders in Appendix section \ref{appendixa22} and use various different measures of neighborhood racial composition in Appendix section \ref{appendixa26}.  Results are consistent with the conclusions presented above.}  

Figure \ref{fig:time_series} plots the evolution of the racial disparity from 2004-2013 and shows that while the racial disparity is present over our whole sample period, it increases dramatically during the Great Recession. This pattern could be taken as evidence that majority-black neighborhoods were less able to mitigate the negative shocks associated with the recession or that the recession disproportionately impacted these neighborhoods.  According to estimates provided by the United States Census Bureau in 2016, the typical black household has a net worth of \$12,920, while that of a typical white household is \$114,700 - this is a \$101,780 difference in wealth that could have important implications for a household's ability to mitigate adverse income shocks. About \$35,000 of this wealth gap is not driven by home equity. By translating this wealth gap into a difference in annual income and using our estimates of the relationship between income and the likelihood of receiving a judgment on your credit score, we calculate that a wealth gap of this size would explain the racial judgment gap.

%% file: Sections/conclusion.tex
\section{Conclusion} \label{sec:conclusion}

We find that Black or Hispanic individuals are 39\% more likely to accumulate debt collection judgment on their credit report, even after controlling for credit score and other credit characteristics.  The debt collection gap does not appear to be driven by states with high levels of racial bias, but does correlate with areas with a high number of payday lenders, many income-less households, and lower shares of tertiary education. We use a second dataset containing zip codes in Missouri to explore potential explanations for the disparity in debt collection judgments and conclude that the racial disparity in judgments is unlikely to be driven by strategic decisions of creditors based on the likelihood of attorney representation or the outcomes of previous cases.  

Given that differences in state-level variation in taste-based discrimination, the likelihood of attorney representation, and previous case outcomes do not appear to be driving our results, we conclude that the persisting disparities in judgment rates may be indicative of systemic racism resulting in financial hardships that are not reflected in the credit report data. Keil and Waldman (2015) quote Lance LeCombs, the Metropolitan St. Louis Sewer District's spokesman, who claims his company has no demographic data on its customers and treats them all the same. He said the racial disparity in its suits was the result of ``broader ills in our community that are outside of our scope and exceed our abilities and authority to do anything about." One such broader ill is the average difference in wealth across black and white households. Estimates provided by the United States Census Bureau in 2016 suggest that the average black household has a net worth of \$12,920, contrasting sharply with the typical white household's net worth of \$114,700 - a staggering wealth gap of \$101,780. This significant disparity in wealth could critically affect a household's ability to weather negative income shocks. Back of the envelope calculations suggest that closing the wealth gap would help to close the judgment gap.  

As the number of debt collection cases rises, it is crucial to identify the extent to which racial disparities exist and how they enter the debt collection system. Future research should explore policies meant to provide more protection to consumers and their impact on the racial disparity in debt collection judgments. For example, such reforms could require debt-buying companies to prove they own the debt before they can sue a debtor, preventing companies from winning judgments when the statute of limitations has expired on a debt\footnote{In most states, the law currently requires defendants to know that the statute of limitations has expired, and raise it as a defense in court.}, or require collection attorneys to prove they have a legal right to collect attorney fees and provide an itemized list of their work on the case to win an attorney's fee through a default judgment\footnote{Currently when companies sue, they often request such fees, which are usually granted and passed on to the debtor as part of the judgment. For example, in Missouri, the fees are typically set at 15 percent of the debt owed, even though attorneys may spend only a few minutes on a suit.}. When states provide legal protections for debtors, such as allowing those with children to keep more of their pay under a head of family exemption, the burden is typically on the debtor to assert these protections. Another policy reform could require a clear notice that these are provided to debtors.

%% file: Sections/figures.tex
\newpage

\section*{Figures}

\begin{figure}[!h]    
\footnotesize
\begin{center}
	\includegraphics[scale=0.6]{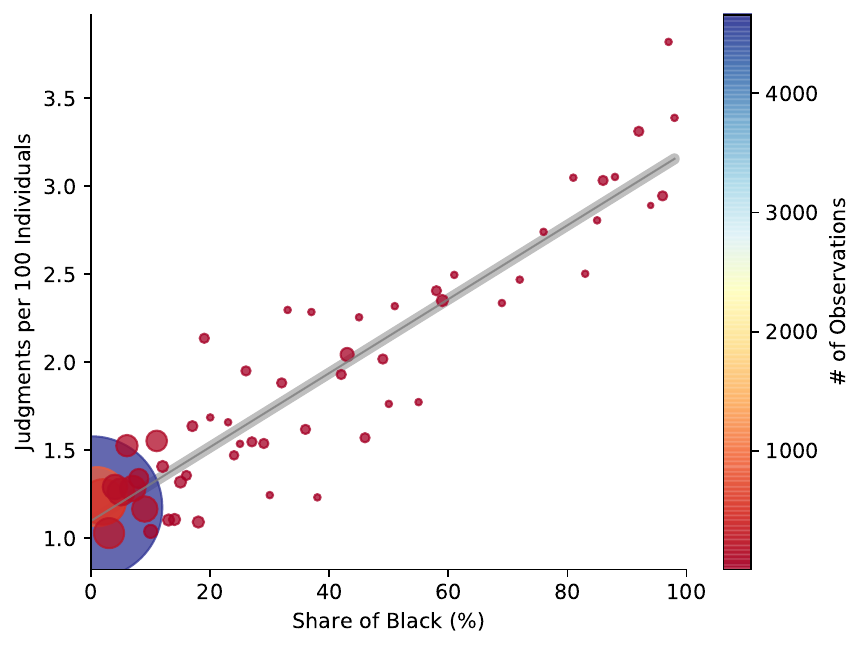}
	\caption{Judgments and Demographic Composition}\label{fig:demographics}
\end{center}
\begin{flushleft}
	\scriptsize{Notes: This figure plots a linear regression that examines the correlation between black population and the judgment rate. The sample includes Missouri zip codes in the common support sample.  Zip codes were sorted into 100 groups based on their black population share. The graph displays each group's average black population share against their average judgment rate. The bubble size corresponds to the data quantity in each group, and the regression line shows the weighted relationship between the judgment rate and the black population share.}
\end{flushleft}
\end{figure}

\begin{figure}[!h]
\footnotesize
\begin{center}
	\begin{subfigure}{0.45\textwidth}  % Adjusted width to make two figures fit side by side
		\centering
		\includegraphics[width=\textwidth]{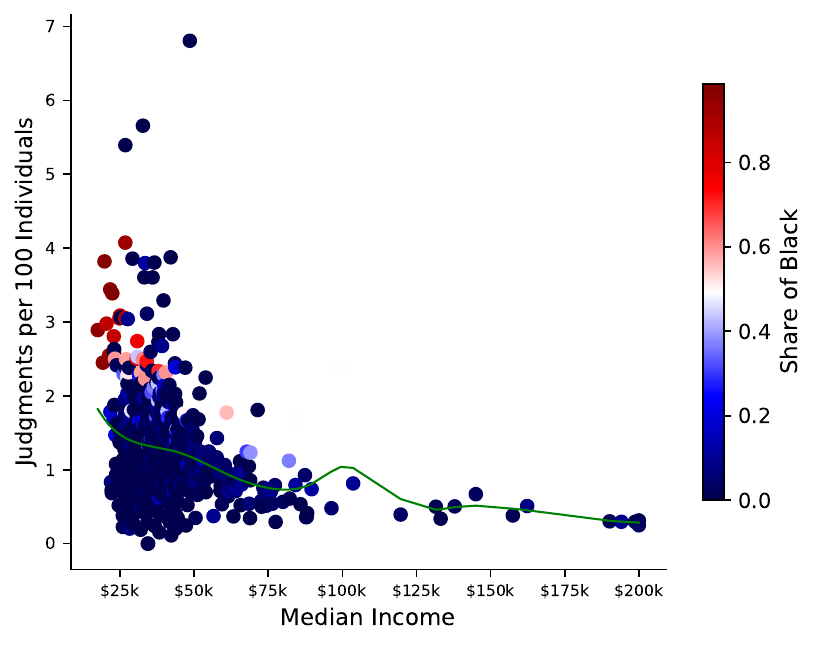}
		\caption{Median Income and Judgment Rate}
	\end{subfigure}
    \hfill  % Added horizontal fill to ensure some space between the figures
	\begin{subfigure}{0.45\textwidth}  % Adjusted width to make two figures fit side by side
		\centering
		\includegraphics[width=\textwidth]{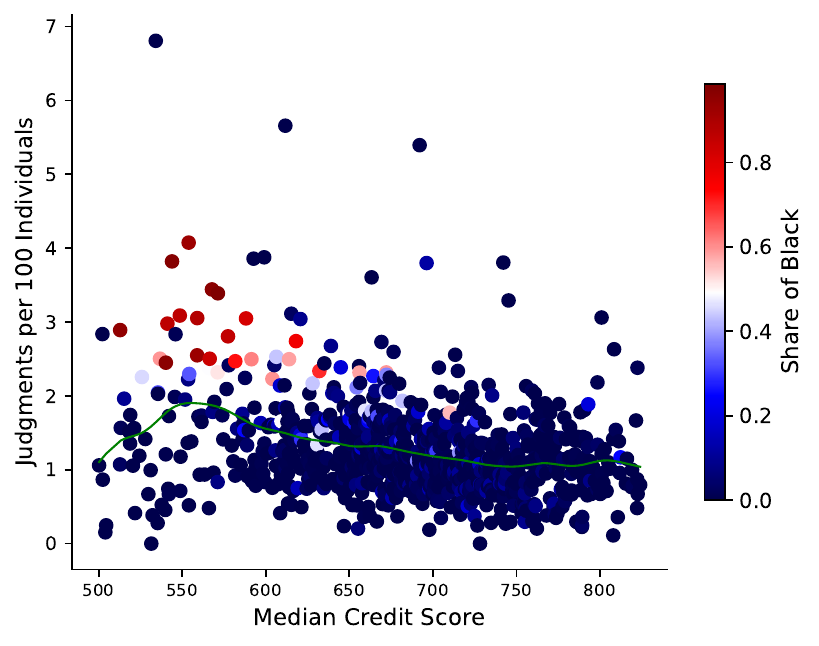}
		\caption{Median Credit Score and Judgment Rate}
	\end{subfigure} \\
\end{center}
	\caption{Income, Credit Scores, and Judgment Rate}\label{fig:cs_vs_income}   
\begin{flushleft}
\scriptsize{Notes: Panel (a) of this figure plots the relationship between median income and the judgment rate and Panel (b) plots the relationship between median credit score and the judgment rate. The sample includes Missouri zip codes in the common support sample. The green line represents the non-parametric locally weighted regression line (LOESS) showing the smoothed fit curve of the data. Income is top coded at \$200K USD to mitigate the impact of outliers.}
\end{flushleft} 
\end{figure}

\begin{figure}[!h]
\footnotesize
	\centering
		\includegraphics[width=.5\textwidth]{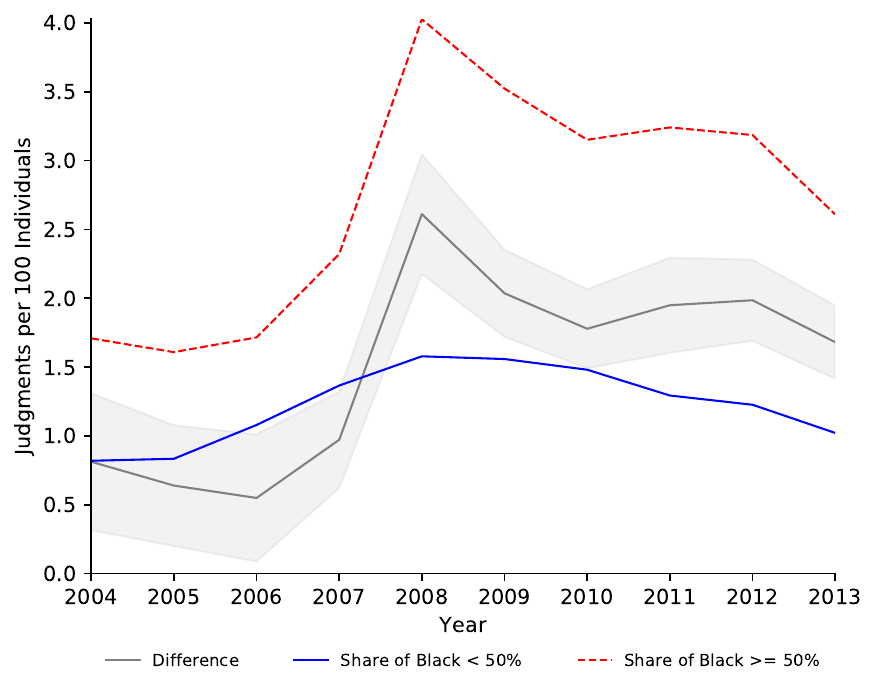}
		\caption{Disparity over Time}\label{fig:time_series}
		\begin{flushleft}
		 \scriptsize{Notes: This figure plots in gray is the estimated coefficient and 95\% confidence interval from a regression which estimates the racial disparity in judgments by year. The sample includes Missouri zip codes in the common support sample. The red dashed lined shows the average number of judgments per 100 individuals for majority black neighborhoods while the blue solid line shows the average number of judgments per 100 individuals for neighborhoods with less than 50\% of black residents.}
		\end{flushleft}
\end{figure}

\newpage

%% file: Sections/tables.tex
\newpage

\input{Tables/tab_1}

\input{Tables/tab_2}

\input{Tables/tab_3}

\input{Tables/Tables_neighborhood/summary_tab1}

\input{Tables/table4_a}

\input{Tables/table_4_bc}

\input{Tables/tab_5}

\input{Tables/Tables_neighborhood/results_debt_portfolio_neighborhood}

\input{Tables/Tables_neighborhood/results_judgment_type_neighborhood}

%% file: Tables/tab_1.tex
\begin{table}[!ht]
\footnotesize
    \centering
    \begin{threeparttable}
        \caption{Descriptive Statistics of Matching Data from Credit Bureau with HMDA}  
        \label{table:1}
        \begin{tabular}{l c c c c c}  \hline \hline
            \multicolumn{6}{l}{\textbf{Panel A: Match rate}} \\ \hline
            & Credit bureau   & Matched to  & Match rate (\%) & & \\
            &  sample &  HMDA &  & & \\
            Home purchase mortgages & 42,296 & 18,611 & 44.00 & &  \\
            Refinance loans & 30,483 & 14,551 & 47.73 & & \\
            All loans & 72,779 & 33,162 & 45.57 & & \\ \hline
            & & & & &  \\[\dimexpr-\normalbaselineskip+2pt]
            \multicolumn{6}{l}{\textbf{Panel B: Home purchase mortgages}} \\ \hline
             & Credit bureau  & Matched to  & Unmatched & \multicolumn{2}{c}{\underline{Matched vs. Unmatched}} \\
             &  sample &  HMDA &  & Norm. diff. & t-stat  \\
            \underline{Match criteria} & & & & & \\
            & & & & & \\[\dimexpr-\normalbaselineskip+2pt]
            Conventional & 0.658 & 0.612 & 0.694 & -0.17 & -17.70 \\
            FHA-insured & 0.278 & 0.320 & 0.244 & 0.17 & 17.54 \\
            VA-guaranteed & 0.065 & 0.068 & 0.063 & 0.02 & 2.20 \\
            Fannie Mae & 0.145 & 0.121 & 0.164 & -0.12 & -12.44 \\
            Freddie Mac & 0.090 & 0.065 & 0.110 & -0.16 & -15.91 \\
            Loan amount & 196,397 & 185,076 & 205,293 & -0.12 & -11.83 \\
            & & & & & \\[\dimexpr-\normalbaselineskip+2pt]
            \underline{Non-match characteristics} & & & & & \\
            & & & & & \\[\dimexpr-\normalbaselineskip+2pt]
            Credit score t-1 & 714 & 709 & 718 & -0.13 & -13.09 \\
            Age & 41 & 39 & 42 & -0.18 & -17.77 \\
            Have mortgage t-1 & 0.120 & 0.084 & 0.148 & -0.20 & -20.31 \\
            Total debt t-1 & 30,034 & 25,899 & 33,283 & -0.11 & -10.87 \\
            Past due debt t-1 & 107 & 103 & 110 & -0.00 & -0.29 \\ \hline
            Observations &  42,313 &  18,618 & 23,695 &  &  \\ \hline \hline
            & & & & & \\[\dimexpr-\normalbaselineskip+2pt]
            \multicolumn{6}{l}{\textbf{Panel C: Refinance loans}} \\  \hline
             & Credit bureau  & Matched to & Unmatched & \multicolumn{2}{c}{\underline{Matched vs. Unmatched}} \\
              &  sample & HMDA &  &  Norm. diff & t-stat   \\
             \underline{Match criteria} & & & & & \\
             & & & & & \\[\dimexpr-\normalbaselineskip+2pt]
            Conventional & 0.864 & 0.850 & 0.877 & -0.08 & -6.91 \\
            FHA-insured & 0.095 & 0.106 & 0.085 & 0.07 & 6.03 \\
            VA-guaranteed & 0.041 & 0.044 & 0.037 & 0.03 & 3.02 \\
            Fannie Mae & 0.132 & 0.135 & 0.129 & 0.02 & 1.61 \\
            Freddie Mac & 0.087 & 0.081 & 0.092 & -0.04 & -3.37 \\
            Loan amount & 219,642 & 204,014 & 233,916 & -0.17 & -14.99 \\
            & & & & & \\[\dimexpr-\normalbaselineskip+2pt]
            \underline{Non-match characteristics} & & & & & \\
            & & & & & \\[\dimexpr-\normalbaselineskip+2pt]
            Credit score t-1 & 732 & 731 & 733 & -0.03 & -2.61 \\
            Age & 48 & 49 & 48 & 0.05 & 3.99 \\
            Have mortgage t-1 & 1.000 & 1.000 & 1.000 & & \\
            Total debt t-1 & 236,565 & 221,709 & 250,134 & -0.15 & -13.41 \\
            Past due debt t-1 & 992 & 557 & 1,389 & -0.05 & -4.24 \\ \hline 
            Observations    & 30,495 & 14,554 & 15,941 &  & \\
                \hline \hline
\end{tabular}
\begin{tablenotes}
\scriptsize
\item Notes: This table provides an overview of the credit bureau and the Home Mortgage Disclosure Act (HMDA) data matching. The initial sample of credit bureau mortgages comprises both home purchase mortgages and refinance loans issued between 2007 and 2017. Loan applicants must apply individually (excluding joint applications), reside within a metropolitan statistical area after loan origination, and have the mortgage as their sole first-lien mortgage. The matching process is described in Section 3.1. Panel A illustrates the success rate of the matching methodology. Panel B summarizes loan and borrower features for home purchase mortgages in the credit bureau data, the subset successfully matched to HMDA, and the unmatched loans. The final two columns display the normalized difference and the outcome of a t-test comparing the mean of the matched sample to the mean of the unmatched sample. Panel C offers comparable summary statistics for refinance loans. 
\end{tablenotes}
\end{threeparttable}
\end{table}

%% file: Tables/tab_2.tex
\begin{table}[!ht]
\centering
\def\sym#1{\ifmmode^{#1}\else\(^{#1}\)\fi}
\footnotesize
\begin{threeparttable}
\caption{Does borrower race affect the Credit Bureau/HMDA match?}
\label{table:2}
\begin{tabular}{lccc}
\hline \hline
& \multicolumn{1}{c}{Full sample} & \multicolumn{1}{c}{Home purchase mortgages} & \multicolumn{1}{c}{Refinance loans} \\
& (1) & (2) & (3) \\
\hline
& & & \\[\dimexpr-\normalbaselineskip+2pt]
\textbf{Match criteria} & & & \\
& & & \\[\dimexpr-\normalbaselineskip+2pt]
FHA loan & -0.02*** & -0.0002*** & -0.0009*** \\
& (0.002085) & (0.00002209) & (0.00003341) \\
VA loan & -0.04*** & -0.0004*** & -0.0007*** \\
& (0.002849) & (0.0000323) & (0.00005913) \\
Purchased by Fannie Mae & -0.09*** & -0.0009*** & -0.0011*** \\
& (0.001643) & (0.00001675) & (0.0000215) \\
Purchased by Freddie Mac & -0.10*** & -0.0010*** & -0.0012*** \\
& (0.001827) & (0.0000186) & (0.0000234) \\
log(Loan amount) & -0.02*** & -0.000074*** & 0.000005067 \\
& (0.001294) & (0.00001641) & (0.00002054) \\ 
& & & \\[\dimexpr-\normalbaselineskip+2pt]
\textbf{Non-match characteristics} & & & \\
& & & \\[\dimexpr-\normalbaselineskip+2pt]
Black & -0.0079 & -0.0006 & -0.0006 \\
& (0.006207) & (0.0005) & (0.0006) \\
Hispanic & -0.0049 & -0.0004 & -0.0013* \\
& (0.006786) & (0.0004) & (0.0006) \\
Black X log(Income) & -0.0006 & 0.00004069 & 0.00004955 \\
& (0.000594) & (0.00004432) & (0.00005803) \\
Hispanic X log(Income) & 0.0001 & 0.0000278 & 0.0001* \\
& (0.0006371) & (0.00003903) & (0.00005155) \\
log(Income) & 0.0030*** & -0.0002*** & -0.00009514*** \\
& (0.000243) & (0.00001564) & (0.00001952) \\ \hline 
& & & \\[\dimexpr-\normalbaselineskip+2pt]
Census tract-by-year FE & Yes & Yes & Yes \\ 
& & & \\[\dimexpr-\normalbaselineskip+2pt]
R-squared & 0.01 & 0.0002 & 0.0003 \\
Observations & 25,902,188 & 12,713,700 & 12,596,574 \\ \hline  \hline
\end{tabular}
\begin{tablenotes}
\scriptsize
\item Notes: This table presents regression analyses investigating the factors influencing the matching of a mortgage from the Home Mortgage Disclosure Act (HMDA) data to a credit bureau record. The sample includes all first-lien home purchase mortgages and refinance loans for owner-occupied properties in metropolitan statistical areas between 2007 and 2017 with reported income. Only single-applicant loans are considered, excluding joint applications. The dependent variable is a binary indicator representing a successful HMDA mortgage match to a credit bureau record, while the independent variables are loan and borrower attributes from the HMDA data. Columns 1, 2, and 3 display the results for the entire sample, home purchase mortgages, and refinance loans, respectively. The coefficients are reported in percentage point units. Standard errors are clustered by census tract-year. \sym{*} \(p<0.10\), \sym{**} \(p<0.05\), \sym{***} \(p<0.01\).
\end{tablenotes}
\end{threeparttable}
\end{table}

%% file: Tables/tab_3.tex
\begin{table}[!ht]
\footnotesize
\centering
\begin{threeparttable}
\caption{Summary Statistics for the Credit Bureau/HMDA Matched Panel}\label{table:3}
\begin{tabular}{lcccccc}
\hline \hline
& & & & & & \\[\dimexpr-\normalbaselineskip+3pt]
 & \multicolumn{1}{p{1.5cm}}{\centering Full credit sample} & \multicolumn{1}{p{1.5cm}}{\centering Matched sample} & \multicolumn{1}{p{1.5cm}}{\centering White} & \multicolumn{1}{p{1.5cm}}{\centering Black} & \multicolumn{1}{p{1.5cm}}{\centering Hispanic}& \multicolumn{1}{p{1.5cm}}{\centering Asian}\\
\hline
& & & & & & \\[\dimexpr-\normalbaselineskip+3pt]
Credit score (t-1) & 678.52 & 712.03 & 718.29 & 668.04 & 686.65 & 733.54 \\
 & (113.98) & (87.67) & (85.52) & (94.40) & (88.29) & (76.98)\\
 & & & & & & \\[\dimexpr-\normalbaselineskip+3pt]
Household income & 77.76 & 94.98 & 97.17 & 78.80 & 81.35 & 112.32 \\
 & (50.31) & (52.79) & (53.09) & (42.10) & (46.69) & (61.57)\\
 & & & & & & \\[\dimexpr-\normalbaselineskip+3pt]
Age & 50.29 & 46.04 & 46.19 & 46.99 & 44.55 & 45.19 \\
 & (19.14) & (14.16) & (14.33) & (14.06) & (13.49) & (12.77)\\
 & & & & & & \\[\dimexpr-\normalbaselineskip+3pt]
Have mortgage (t-1) & 0.27 & 0.70 & 0.72 & 0.65 & 0.63 & 0.66 \\
 & (0.44) & (0.46) & (0.45) & (0.48) & (0.48) & (0.47)\\
 & & & & & & \\[\dimexpr-\normalbaselineskip+3pt]
Total debt (t-1) & 62,464 & 155,939 & 157,842 & 127,641 & 133,802 & 207,858 \\
 & (142,691) & (176,230) & (178,632) & (123,868) & (150,118) & (226,610)\\
 & & & & & & \\[\dimexpr-\normalbaselineskip+3pt]
Past due debt (t-1) & 2,040.78 & 2,606.64 & 2,157.72 & 5,615.54 & 3,731.79 & 2,441.48 \\
 & (23,461.18) & (24,735.67) & (21,408.84) & (32,312.05) & (29,969.48) & (40,222.92)\\
 & & & & & & \\[\dimexpr-\normalbaselineskip+3pt]
New judgment outstanding & 0.1698 &  0.1428 &  0.1308 &  0.3189 &  0.1416 &  0.0583  \\
 & (4.1171) & (3.7762) & (3.6143) & (5.6378) & (3.7602) & (2.4136) \\
  & & & & & & \\[\dimexpr-\normalbaselineskip+3pt]
New judgment petitioned & 0.0827 & 0.0702 & 0.0648 & 0.1728 & 0.0527 & 0.0276  \\
 &  (2.8745) & (2.6478) & (2.5444) & (4.1535) & (2.2960) & (1.6605) \\
  & & & & & & \\[\dimexpr-\normalbaselineskip+3pt]
New judgment satisfied & 0.0283 & 0.0304 & 0.0278 & 0.0741 & 0.0243 & 0.0138  \\
 & (1.6811) & (1.7435) & (1.6682) & (2.7205) & (1.5601) & (1.1742) \\
\hline
Observations  & 769,995  & 28,361  & 21,658  & 2,276  & 2,789 & 1,638 \\ \hline \hline
\end{tabular}
\begin{tablenotes}
\item \scriptsize{This table provides descriptive statistics for the Credit Bureau/HMDA matched panel. Column 1 outlines the sample means and standard deviations for the entire credit bureau dataset; column 2 portrays these statistics for the Credit Bureau/HMDA matched panel; and columns 3–6 display these statistics for the White, Black, Hispanic, Asian borrowers within the matched dataset, respectively. The sample spans 2013Q1 to 2017Q2.  }
\end{tablenotes}
\end{threeparttable}
\end{table}

%% file: Tables/Tables_neighborhood/summary_tab1.tex
\begin{table}[!h]\centering
  \begin{threeparttable}
 \footnotesize 
\def\sym#1{\ifmmode^{#1}\else\(^{#1}\)\fi}
\caption{Summary Statistics for Neighborhood Level Judgment Data\label{tab:summary_statistics}}
\begin{tabular}{l*{3}{cc}}
\hline\hline
                    &\multicolumn{1}{c}{Black}&\multicolumn{1}{c}{Non-Black}&\multicolumn{1}{c}{Difference}\\
\hline
& & & \\[\dimexpr-\normalbaselineskip+2pt]
\textbf{Panel A: Judgments and Financial Characteristics } & & & \\
 \textcolor{white}{...}Judgments per 100 People       &        2.71&        1.26&       -1.46\sym{***}\\
                    &      (1.29)&      (0.75)&                     \\
 \textcolor{white}{...}Share of Default Judgments            &        0.45&        0.37&       -0.08\sym{***}\\
                    &      (0.07)&      (0.13)&                     \\
 \textcolor{white}{...}Share of Consent Judgments            &        0.16&        0.17&       0.01\sym{*}\\
                    &      (0.07)&      (0.11)&                     \\
 \textcolor{white}{...}Share of Dismissed Judgments           &        0.11&        0.15&        0.04\sym{***}\\
                    &      (0.11)&      (0.13)&                     \\
 \textcolor{white}{...}Share of  Contested Judgments       &        0.05&        0.05&       -0.01\sym{*}  \\
                    &      (0.04)&      (0.07)&                     \\
 \textcolor{white}{...}Share of  Settled Cases              &        0.19&        0.2&        0.01\sym{*}\\
                    &      (0.09)&      (0.13)&                     \\
 \textcolor{white}{...}Share w/ Attorney            &        0.04&        0.10&        0.06\sym{***}\\
                    &      (0.02)&      (0.08)&                     \\
 \textcolor{white}{...}Mean Household Income from IRS (in 000s) &       29.21&       38.68&        9.47\sym{***}\\
                    &      (9.75)&      (9.88)&                     \\
 \textcolor{white}{...}Median Credit Score &    587.59&      686.36&       98.77\sym{***}\\
                  &     (49.83)&     (72.68)&                     \\
 \textcolor{white}{...}90+ DPD Debt Balances             &     3237.66&     1405.79&    -1831.87\sym{***}\\
 &   (2629.86)&   (3768.89)&                     \\
\textbf{Panel B: Neighborhood Characteristics} & & & \\ 
 \textcolor{white}{...}GINI Index &        0.46&        0.41&       -0.05\sym{***}\\
                    &      (0.05)&      (0.05)&                     \\

 \textcolor{white}{...}Unemployment Rate   &        0.12&        0.05&       -0.06\sym{***}\\
                    &      (0.03)&      (0.03)&                     \\
 \textcolor{white}{...}Divorce Rate&        0.13&        0.12&       -0.01\sym{***}\\
                    &      (0.02)&      (0.04)&                     \\
 \textcolor{white}{...}Fraction with Bachelors Degree&        0.16&        0.16&       -0.00         \\
                    &      (0.10)&      (0.10)&                     \\
\textcolor{white}{...}Median Gross Rent  &        0.76&        0.62&       -0.14\sym{***}\\
                    &      (0.14)&      (0.16)&                     \\
\textcolor{white}{...}Home Ownership Rate &        0.50&        0.74&        0.25\sym{***}\\
                    &      (0.16)&      (0.12)&                     \\
 %\textcolor{white}{...}Banks (5 miles)     &       88.10&       11.91&      -76.19\sym{***}\\
 %                   &     (39.71)&     (30.19)&                     \\
 %\textcolor{white}{...}Payday Lenders (5 miles)&       30.64&        3.58&      -27.07\sym{***}\\
 %                   &      (9.23)&      (8.17)&                     \\
 %\textcolor{white}{...}Zillow: Single Family House (in 000s) &       90.28&      113.32&       23.04\sym{***}\\
 %                   &     (54.39)&     (43.23)&                     \\
 %\textcolor{white}{...}HMDA: Share Rejected, Black - White&        0.19&        0.14&       -0.05\sym{***}\\
 %                   &      (0.11)&      (0.31)&                     \\
\hline
Zip codes       &         247&        6640&            \\
Average Population    &    18451.03&     7036.03&   \\
\hline\hline
\end{tabular}

   \begin{tablenotes}
     \scriptsize
    \item Notes:  This table presents summary statistics for the Missouri neighborhood level data from 2004-2013.  The sample includes only observations on the common support.  Column 1 displays the sample means and standard deviations for neighborhoods in which at least 50\% of the population is black and column 2 displays sample means and standard deviations for neighborhoods in which less than 50\% is black.  The third column reports the difference in means across the two samples. \sym{*} \(p<0.10\), \sym{**} \(p<0.05\), \sym{***} \(p<0.01\)
    \end{tablenotes}
\end{threeparttable}
\end{table}

%% file: Tables/table4_a.tex
\begin{sidewaystable}[!ht]
\footnotesize
\centering
\begin{threeparttable}
\caption{Race and New Outstanding Judgments}\label{table:4a}
\def\sym#1{\ifmmode^{#1}\else\(^{#1}\)\fi}
\begin{tabular*}{\hsize}{@{\hskip\tabcolsep\extracolsep\fill}l*{6}{c}}
\hline\hline
                    &\multicolumn{1}{c}{(1)}&\multicolumn{1}{c}{(2)}&\multicolumn{1}{c}{(3)}&\multicolumn{1}{c}{(4)}&\multicolumn{1}{c}{(5)}&\multicolumn{1}{c}{(6)}\\
                    &\multicolumn{6}{c}{Outstanding Judgment}\\

\hline
& & & & & & \\[\dimexpr-\normalbaselineskip+3pt]
Female         &      0.0114         &      0.0037         &      0.0086         &      0.0111         &      0.0397         &     -0.0059         \\
                    &    (0.0111)         &    (0.0138)         &    (0.0136)         &    (0.0139)         &    (0.0571)         &    (0.0067)         \\
& & & & & & \\[\dimexpr-\normalbaselineskip+3pt]
Minority             &      0.0974\sym{***}&      0.1156\sym{***}&      0.0496\sym{**} &      0.0470\sym{*}  &      0.1373\sym{*}  &      0.0194         \\
                    &    (0.0202)         &    (0.0246)         &    (0.0238)         &    (0.0245)         &    (0.0756)         &    (0.0122)         \\
& & & & & & \\[\dimexpr-\normalbaselineskip+3pt]
Age                         &     -0.0013\sym{***}&     -0.0018\sym{***}&      0.0008\sym{*}  &      0.0004         &      0.0025         &      0.0003         \\
                    &    (0.0003)         &    (0.0004)         &    (0.0004)         &    (0.0005)         &    (0.0020)         &    (0.0002)         \\
& & & & & & \\[\dimexpr-\normalbaselineskip+3pt]
Credit Score$_{t-1}$  &                     &                     &     -0.0024\sym{***}&     -0.0026\sym{***}&     -0.0044\sym{***}&     -0.0004\sym{***}\\
                    &                     &                     &    (0.0002)         &    (0.0002)         &    (0.0005)         &    (0.0001)         \\
& & & & & & \\[\dimexpr-\normalbaselineskip+3pt]
Past Due Debt$_{t-1}$      &                     &                     &      0.0019\sym{***}&      0.0022\sym{***}&      0.0011         &      0.0019         \\
                    &                     &                     &    (0.0006)         &    (0.0007)         &    (0.0008)         &    (0.0020)         \\
& & & & & & \\[\dimexpr-\normalbaselineskip+3pt]
Total Debt$_{t-1}$          &                     &                     &     -0.0002\sym{***}&     -0.0004\sym{***}&     -0.0006\sym{***}&     -0.0000         \\
                    &                     &                     &    (0.0000)         &    (0.0001)         &    (0.0002)         &    (0.0000)         \\
& & & & & & \\[\dimexpr-\normalbaselineskip+3pt]
Outstanding Judgment$_{t-1}$   &                     &                     &      0.0216         &     -0.0020         &     -0.8370\sym{***}&     -0.2061         \\
                    &                     &                     &    (0.0991)         &    (0.1008)         &    (0.1410)         &    (0.1615)         \\
& & & & & & \\[\dimexpr-\normalbaselineskip+3pt]
Constant            &      0.1831\sym{***}&      0.2067\sym{***}&      1.8568\sym{***}&      1.9283\sym{***}&      3.0859\sym{***}&      0.3312\sym{***}\\
                    &    (0.0193)         &    (0.0219)         &    (0.1186)         &    (0.1270)         &    (0.3127)         &    (0.0621)         \\
\hline
Borrowers  &  All    &   All  &  All  &   All  &   Subprime  &  Prime  \\
State X Quarter Fixed Effects&                     &           X         &           X         &           X         &           X         &           X         \\
ZIP Fixed Effects   &                     &           X         &           X         &           X         &           X         &           X         \\
Observations        & 469020         & 468754         & 468753         & 445530         & 122640         & 345323         \\
Mean of DV (Minority = 0)           &      0.1258         &      0.1258         &      0.1258         &      0.1258         &      0.4817         &      0.0187         \\
$R^2$               &      0.0001         &      0.0362         &      0.0385         &      0.0399         &      0.0744         &      0.0339         \\
\hline\hline
\end{tabular*}
\begin{tablenotes}
\scriptsize
\item Notes:  This table reports results from a regression of judgments on race, individual characteristics, zip code fixed effects, and state-by-quarter fixed effects. The data comprises credit bureau records from 2013–2017, which have been merged with Home Mortgage Disclosure Act records. The dependent variable is a binary indicator denoting having a new judgment outstanding. The sample is quarterly credit report data. Column 1 includes no fixed effects and controls for age, gender, and minority status. Column 2 adds fixed effects, and Column 3 includes credit report controls.  Column 4 adds imputed credit bureau income, as well as HMDA income to the controls used in Column 3. Columns 5 and 6 focus specifically on applicants with subprime and prime credit scores, respectively, using the specification from Column 3. Reported coefficients are in percentage point units. Standard errors are clustered by state-quarter. \sym{*} \(p<0.10\), \sym{**} \(p<0.05\), \sym{***} \(p<0.01\). 
\end{tablenotes}
\end{threeparttable}
\end{sidewaystable}

%% file: Tables/table_4_bc.tex
\begin{sidewaystable}[!ht]
\footnotesize
\centering
\begin{threeparttable}
\caption{Race and Judgments Petitioned or Satisfied}\label{table:4bc}
\def\sym#1{\ifmmode^{#1}\else\(^{#1}\)\fi}
\begin{tabular*}{\hsize}{@{\hskip\tabcolsep\extracolsep\fill}l*{6}{c}}
\hline\hline
                    &\multicolumn{1}{c}{(1)}&\multicolumn{1}{c}{(2)}&\multicolumn{1}{c}{(3)}&\multicolumn{1}{c}{(4)}&\multicolumn{1}{c}{(5)}&\multicolumn{1}{c}{(6)}\\
\hline
\textbf{Panel A:  New Petitioned Judgment}  & & & & & & \\
& & & & & & \\[\dimexpr-\normalbaselineskip+3pt]
Minority             &      0.0464\sym{***}&      0.0555\sym{***}&      0.0220         &      0.0192         &      0.0575         &      0.0031         \\
                    &    (0.0132)         &    (0.0180)         &    (0.0178)         &    (0.0186)         &    (0.0572)         &    (0.0077)         \\
& & & & & & \\[\dimexpr-\normalbaselineskip+3pt]
\hline
\textbf{Panel B:  New Satisfied Judgment}  & & & & & & \\
& & & & & & \\[\dimexpr-\normalbaselineskip+3pt]
Minority      &      0.0222\sym{**} &      0.0282\sym{***}&      0.0120         &      0.0068         &      0.0152         &      0.0062         \\
                    &    (0.0091)         &    (0.0106)         &    (0.0103)         &    (0.0106)         &    (0.0253)         &    (0.0090)         \\ \hline
Borrowers  &  All    &   All  &  All  &   All  &   Subprime  &  Prime  \\
Credit Controls  &      &   &  X  &   X  &   X  &  X  \\
State X Quarter Fixed Effects&                     &           X         &           X         &           X         &           X         &           X         \\
ZIP Fixed Effects   &                     &           X         &           X         &           X         &           X         &           X         \\
Observations        & 469020         & 468754         & 468753         & 445530         & 122640         & 345323         \\
Mean of DV (Minority = 0)           &      0.0615         &      0.0615         &      0.0615         &      0.0615         &      0.2370         &      0.0087         \\
$R^2$               &      0.0001         &      0.0343         &      0.0356         &      0.0366         &      0.0707         &      0.0277         \\
\hline\hline
\end{tabular*}
\begin{tablenotes}
\scriptsize
\item Notes:  This table tests a regression of judgments on race, individual characteristics, zip code fixed effects, and state-by-quarter fixed effects. The data comprises credit bureau records from 2013–2017, which have been merged with Home Mortgage Disclosure Act records. The dependent variable in Panel A is a binary indicator denoting having a new petitioned judgment, and the dependent variable in Panel B is a binary variable denoting having a new satisfied judgment. The sample is quarterly credit report data. Column 1 includes no fixed effects and controls for age, gender, and minority status. Column 2 adds fixed effects, while Column 3 includes credit report controls.  Column 4 adds imputed credit bureau income, as well as HMDA income to the controls used in Column 3. Columns 6 and 7 focus specifically on applicants with subprime and prime credit scores, respectively, using the specification from Column 3. Reported coefficients are in percentage point units. Standard errors are clustered by state-quarter. \sym{*} \(p<0.10\), \sym{**} \(p<0.05\), \sym{***} \(p<0.01\). 
\end{tablenotes}
\end{threeparttable}
\end{sidewaystable}

%% file: Tables/tab_5.tex
\begin{sidewaystable}[!ht]
\footnotesize
\centering
\begin{threeparttable}
\caption{Where are racial disparities in debt collection highest?}\label{table:5}
\def\sym#1{\ifmmode^{#1}\else\(^{#1}\)\fi}
\begin{tabular*}{\hsize}{@{\hskip\tabcolsep\extracolsep\fill}l*{5}{c}}
\hline\hline
                    &\multicolumn{1}{c}{(1)}&\multicolumn{1}{c}{(2)}&\multicolumn{1}{c}{(3)}&\multicolumn{1}{c}{(4)}&\multicolumn{1}{c}{(5)}\\
                    &\multicolumn{5}{c}{Judgment Outstanding}\\

\hline
& & & & & \\[\dimexpr-\normalbaselineskip+3pt]
Minority            &      0.0496\sym{**} &      0.0472\sym{*}  &      0.0883\sym{***}&      0.0213         &      0.0674\sym{**} \\
                    &    (0.0238)         &    (0.0271)         &    (0.0314)         &    (0.0244)         &    (0.0273)         \\
& & & & & \\[\dimexpr-\normalbaselineskip+3pt]
Minority $\times$ High Racial Bias &                     &      0.0107         &                     &                     &                     \\
                    &                     &    (0.0594)         &                     &                     &                     \\
& & & & & \\[\dimexpr-\normalbaselineskip+3pt]
Minority $\times$ Low Non-Bank Financing&                     &                     &     -0.1096\sym{**} &                     &                     \\
                    &                     &                     &    (0.0468)         &                     &                     \\
& & & & & \\[\dimexpr-\normalbaselineskip+3pt]
Minority $\times$ High Households without Income &                     &                     &                     &      0.1455\sym{**} &                     \\
                    &                     &                     &                     &    (0.0719)         &                     \\
& & & & & \\[\dimexpr-\normalbaselineskip+3pt]
Minority $\times$ High Tertiary Education&                     &                     &                     &                     &     -0.0888\sym{*}  \\
                    &                     &                     &                     &                     &    (0.0477)         \\
\hline
Credit Report Controls&           X         &           X         &           X         &           X         &           X         \\
State X Quarter Fixed Effects&           X         &           X         &           X         &           X         &           X         \\
ZIP Fixed Effects   &           X         &           X         &           X         &           X         &           X         \\
Observations        & 468753        & 468641        & 463959        & 464026        & 464076        \\
Mean of DV (Minority=0)          &      0.1258         &      0.1258         &      0.1258         &      0.1258         &      0.1258         \\
$R^2$               &      0.0385         &      0.0385         &      0.0385         &      0.0385         &      0.0385         \\
\hline\hline
\end{tabular*}
\begin{tablenotes}
\scriptsize
\item Notes:  This table presents results from regression analyses examining debt collection judgments as a function of race, individual attributes, zip code fixed effects, and state-by-quarter fixed effects. The individual-level data are sourced from a combined dataset of credit bureau records and Home Mortgage Disclosure Act records. This dataset incorporates credit bureau records spanning the years from 2013 through 2018. The dependent variable is a binary indicator reflecting new outstanding debt collection judgment. Each specification controls for age, gender, minority status, state by quarter fixed effects, zip code fixed effects, and credit report controls. The reported coefficients are in percentage points. Standard errors are calculated using clustering by state-quarter. \sym{*} \(p<0.10\), \sym{**} \(p<0.05\), \sym{***} \(p<0.01\).
\end{tablenotes}
\end{threeparttable}
\end{sidewaystable}

%% file: Tables/Tables_neighborhood/results_debt_portfolio_neighborhood.tex
\begin{table}[htbp] \centering
\begin{threeparttable}
\footnotesize
\def\sym#1{\ifmmode^{#1}\else\(^{#1}\)\fi}
\caption{Judgments and Debt Portfolios\label{tab:debt_composition}}
\begin{tabular}{l*{5}{c}}
\hline\hline
                    &\multicolumn{1}{c}{(1)}&\multicolumn{1}{c}{(2)}&\multicolumn{1}{c}{(3)}&\multicolumn{1}{c}{(4)}&\multicolumn{1}{c}{(5)}\\
                       &\multicolumn{5}{c}{Judgment Rate}\\
\hline
& & & & & \\[\dimexpr-\normalbaselineskip+2pt]
Black Majority: ZIP &      0.6910\sym{***}&      0.7134\sym{***}&      0.7070\sym{***}&      0.7095\sym{***}&      0.6962\sym{***}\\
                    &    (0.1616)         &    (0.1657)         &    (0.1650)         &    (0.1579)         &    (0.1511)         \\
\hline
& & & & &  \\[\dimexpr-\normalbaselineskip+2pt]
Debt Levels         &           X         &                     &                     &                     &           X         \\
Monthly Payment and Utilization&                     &           X         &                     &                     &           X         \\
Debt Composition    &                     &                     &           X         &                     &           X         \\
Delinquency/Bankruptcy/Collections&                     &                     &                     &           X         &           X         \\
\hline
& & & & & \\[\dimexpr-\normalbaselineskip+2pt]
Wild Cluster Bootstrap p-value&       .0012         &       .0009         &       .0009         &       .0005         &       .0004         \\
Observations        &   7019      &   7019      &   7019      &   7019      &   7019      \\
$R^2$               &      0.6877         &      0.6860         &      0.6854         &      0.6876         &      0.6926         \\
\hline\hline
\end{tabular}
\begin{tablenotes}
\scriptsize
\item Notes:  This table presents results of a regression of the judgment rate on neighborhood racial composition and other neighborhood-level characteristics.  The sample includes Missouri zip codes in the common support sample.  The dependent variable is judgments per 100 individuals.  Baseline controls include average zip code-level income, the fraction of IRS filings under \$25,000, between \$25,000-50,000, \$50,000-75,000, \$75,000-100,000, and over \$100,000, credit score quintiles, median credit score, total delinquent debt balances, unemployment rate, median house value, the fraction of the population with a college education, the divorce rate, and population density.  Baseline controls, county fixed effects, and year fixed effects are included in all specifications, and all regressions are weighted by population. Robust standard errors clustered at the county-year level are in parentheses. Effective number of clusters: 36.0. Wild Cluster Bootstrap p-values are reported for our main parameter of interest. \sym{*} \(p<0.10\), \sym{**} \(p<0.05\), \sym{***} \(p<0.01\). 
\end{tablenotes}
\end{threeparttable}
\end{table}

%% file: Tables/Tables_neighborhood/results_judgment_type_neighborhood.tex
\begin{sidewaystable}[htbp]\centering
\begin{threeparttable}
\footnotesize
\def\sym#1{\ifmmode^{#1}\else\(^{#1}\)\fi}
\caption{Attorney Representation and Judgment Type\label{tab:judgment_type}}
\begin{tabular}{l*{8}{c}}
\hline\hline
                    &\multicolumn{1}{c}{(1)}&\multicolumn{1}{c}{(2)}&\multicolumn{1}{c}{(3)}&\multicolumn{1}{c}{(4)}&\multicolumn{1}{c}{(5)}&\multicolumn{1}{c}{(6)}&\multicolumn{1}{c}{(7)}&\multicolumn{1}{c}{(8)}\\
                    &\multicolumn{1}{c}{Attorney}&\multicolumn{1}{c}{Judgments}&\multicolumn{1}{c}{Consent}&\multicolumn{1}{c}{Contested}&\multicolumn{1}{c}{Default}&\multicolumn{1}{c}{Dismissed}&\multicolumn{1}{c}{Settle}&\multicolumn{1}{c}{Judgments}\\
\hline
&&&&& & & & \\[\dimexpr-\normalbaselineskip+2pt]
\hline
Black Majority: ZIP &      -0.015\sym{***}&       0.707\sym{***}&       0.003         &       0.001         &       0.013\sym{**} &       0.014\sym{*}  &      -0.022\sym{***} & 0.768\sym{***} \\
                    &     (0.004)             &     (0.165)         &     (0.005)         &     (0.004)         &     (0.006)         &     (0.007)         &     (0.006)     & (0.165)   \\
\hline
&&&&& & & & \\[\dimexpr-\normalbaselineskip+2pt]
Attorney Representation & & X & X&X&X&X&X&X \\
Lagged Case Outcomes & & & & & & & & X \\ \hline 
&&&&& & & & \\[\dimexpr-\normalbaselineskip+2pt]
Wild Cluster Bootstrap p-value&       .0001         &       .0007         &       .0008         &        .601         &        .709         &       .0558         &       .0702         &       .0036         \\
Observations        &    7019       &    7019       &    7019       &    7019       &    7019       &    7019       &    7019       &    7019       \\
$R^2$               &       0.448         &       0.700         &       0.685         &       0.470         &       0.438         &       0.395         &       0.651         &       0.615         \\
\hline\hline
\end{tabular}
\begin{tablenotes}
\scriptsize
\item This table explores the extent to which attorney representation or previous case outcomes can explain the racial gap in judgments.  The sample includes Missouri zip codes in the common support sample. The outcome variable in Column (1) is the share of debt collection court cases where an attorney represented the defendant.  The outcome variable in columns (2) and (8) are judgments per 100 people.  The outcome variables in Columns (3)-(7) are the share of cases resulting in a consent judgment, a contested judgment, a default judgment, the case being dismissed, and the case being settled, respectively.  All regressions are weighted by population. Baseline controls (average zip code-level income, the fraction of IRS filings under \$25,000, between \$25,000-50,000, \$50,000-75,000, \$75,000-100,000, and over \$100,000, credit score quintiles, median credit score, total delinquent debt balances, unemployment rate, median house value, the fraction of the population with a college education, the divorce rate, and population density), county fixed effects, and year fixed effects are included in all specifications. Aside from Column (1), all specifications control for the share of cases with attorney representation.  Robust standard errors clustered at the county-year level are in parentheses. Effective number of clusters: 36.0. Wild Cluster Bootstrap p-values are reported for our main parameter of interest. \sym{*} \(p<0.10\), \sym{**} \(p<0.05\), \sym{***} \(p<0.01\). 
\end{tablenotes}
\end{threeparttable}
\end{sidewaystable}

%% file: Sections/appendix.tex
\appendix
\counterwithin{figure}{section}
\counterwithin{table}{section}

\section{Additional Details and Analysis}\label{sec:appendixa}

\renewcommand{\thefigure}{A\arabic{figure}}
\renewcommand{\thetable}{A\arabic{table}}
\setcounter{figure}{0}
\setcounter{table}{0}

\subsection{Judgment Data}\label{appendixa1}

We obtained our Missouri judgment data from Paul Kiel and Annie  Waldman at ProPublica. The data they shared with us included all debt collection judgments in New Jersey, Missouri, and Cook County, Illinois, from 2008 to 2012.\footnote{The focus in their ProPublica article was Essex County, St. Louis City, St. Louis County, and Cook County due to the cities' high segregation indexes. Due to a peculiarity of the court system database, the Essex County window is slightly different: July 1, 2007, through June 30, 2012.}  Both Missouri and New Jersey have state-wide databases. The Missouri dataset was provided by Missouri's Office of the State Courts Administrator (OSCA) and included all debt collection cases filed in Associate Circuit Court for which OSCA has an electronic record through early 2014.\footnote{The max amount sought in associate circuit courts in Missouri is \$25,000.} For each case in Missouri, the data contained the following information: court (judicial circuit), county, case ID, filing Date, case type, disposition, plaintiff, plaintiff attorney, defendant, defendant date of birth, defendant address, defendant attorney, judgment amount, date of judgment satisfaction, date of first garnishment attempt.\footnote{The judgment amount was determined to be unreliable and is not used throughout this analysis.} Kiel and Waldman added two fields: a standard name for each plaintiff and a plaintiff type. Missouri's court system has some variation among the judicial circuits in how case types are categorized, so Keil and Waldman selected a range of case types that could be reasonably construed as debt collection cases in consultation with OSCA employees. For St. Louis City and County courts, these were: Breach of Contract, Promissory Note, Suit on Account, Contract /Account (Bulk), Misc Associate Civil-Other, Small Claims under \$100, Small Claims over \$100.\footnote{Together, the small claims and ``misc associate'' cases comprised less than four percent of cases.} They limited the dataset to cases that had resulted in a judgment.

\begin{figure}[ht]
\footnotesize
    \centering
    \begin{subfigure}[b]{0.30\textwidth}
        \includegraphics[width=\textwidth]{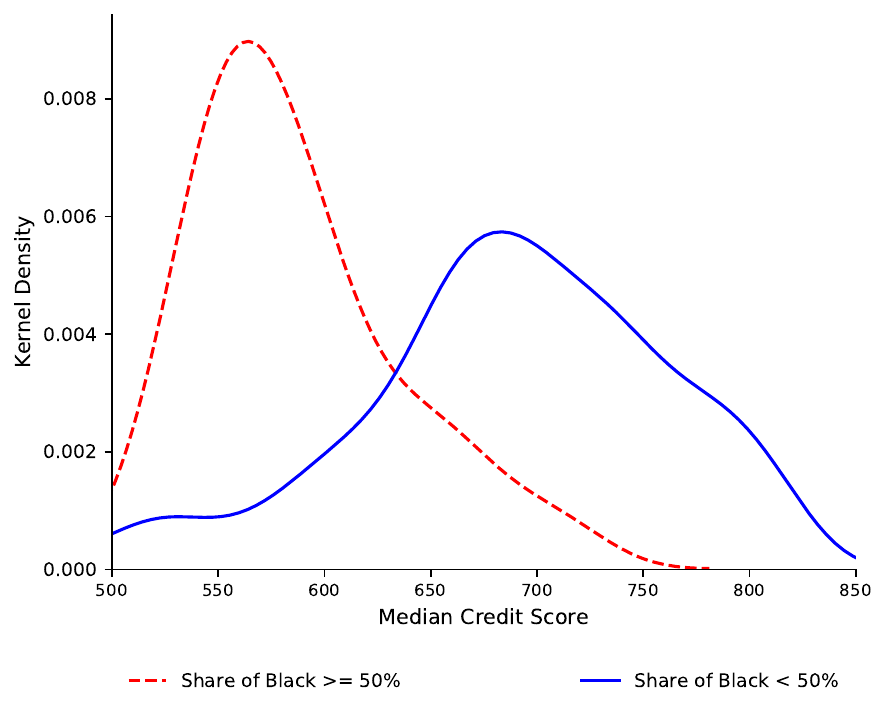}
        \caption{Credit Score}
        \label{fig:cs}
    \end{subfigure}
    \begin{subfigure}[b]{0.30\textwidth}
        \includegraphics[width=\textwidth]{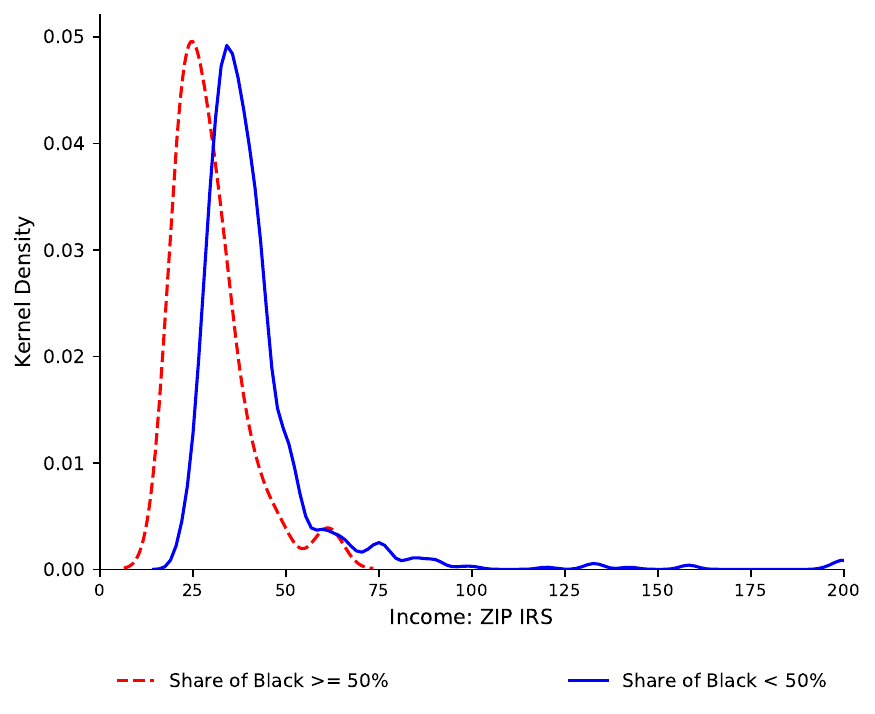}
        \caption{Household Income}
        \label{fig:hi}
    \end{subfigure} \\
    \begin{subfigure}[b]{0.30\textwidth}
        \includegraphics[width=\textwidth]{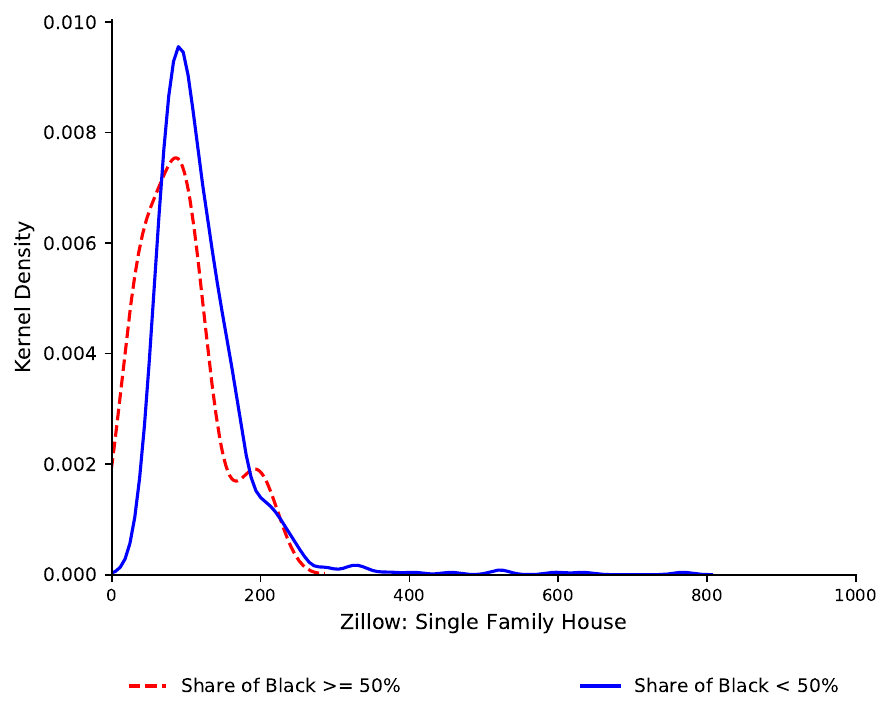}
        \caption{House Value}
        \label{fig:hv}
        \end{subfigure}
     \begin{subfigure}[b]{0.30\textwidth}
        \includegraphics[width=\textwidth]{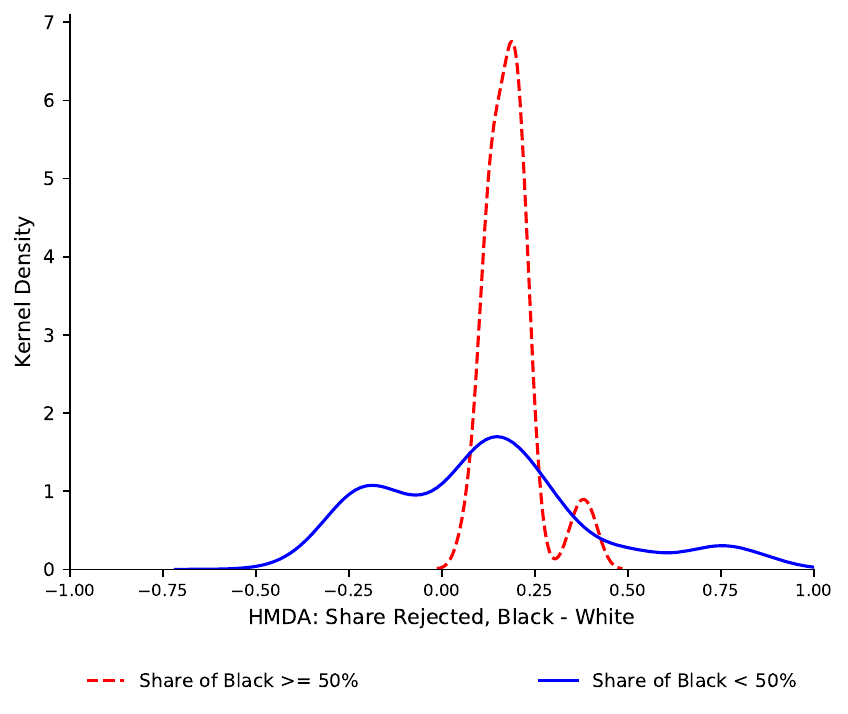}
        \caption{Mortgage Denials}
        \label{fig:hmda}       
    \end{subfigure} \\
        \begin{subfigure}[b]{0.30\textwidth}
        \includegraphics[width=\textwidth]{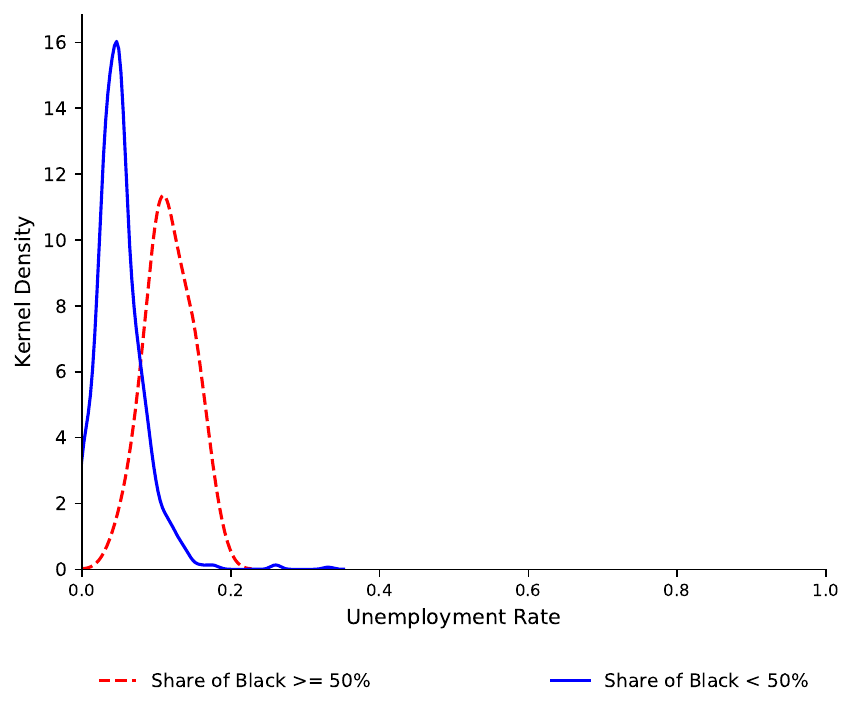}
        \caption{Unemployment Rate}
        \label{fig:ur}
    \end{subfigure}
        \begin{subfigure}[b]{0.30\textwidth}
        \includegraphics[width=\textwidth]{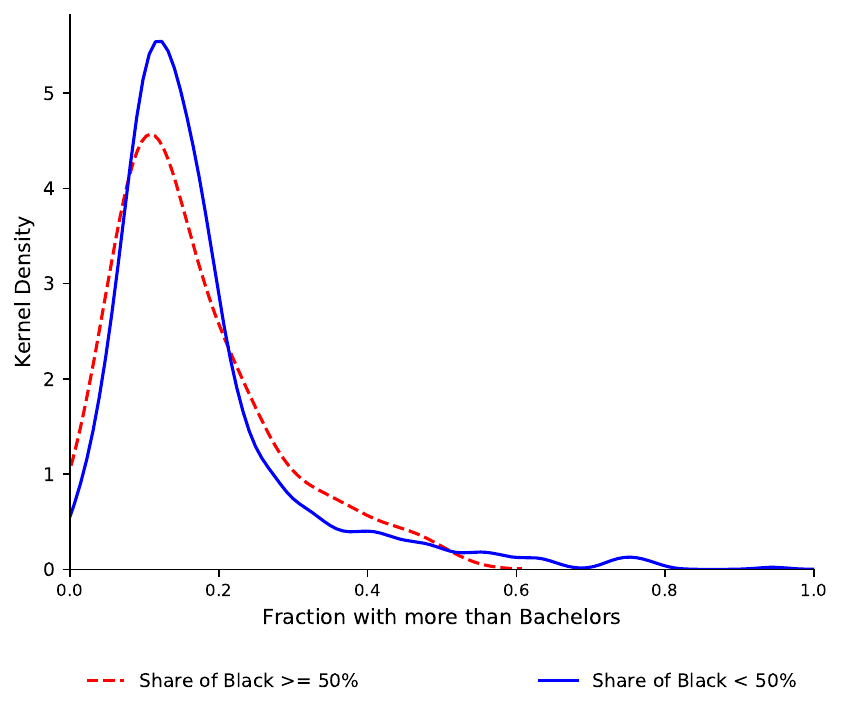}
        \caption{$>=$ Bachelor Degree}
        \label{fig:bd}
    \end{subfigure}
    \caption{Kernel Density Estimates of Selected Covariates}\label{fig:kde}
    \begin{flushleft}
\scriptsize{Notes:  These figures plot the densities of various neighborhood characteristics cross majority black and majority non-black neighborhoods. Data come from the entire Missouri zip code level dataset.}
\end{flushleft} 
\end{figure}

We collapse this data to the zip code level.  As discussed in the Data Section of our paper, our main analysis focuses on Missouri, although we replicate our results with data from New Jersey and Illinois wherever feasible.  Within this data there are important differences in observable characteristics across majority black and non-majority black neighborhoods. Examples of such differences are documented in Figure \ref{fig:kde}. This figure shows kernel density estimates of various covariates used throughout this analysis.  Majority-black neighborhoods tend to have lower median credit scores, lower average incomes, lower house values, higher unemployment rates, and a higher share of divorced individuals. These differences cause concern that there could also be important differences in unobservable characteristics that vary with the racial composition of neighborhoods. We mitigate this concern by limiting our dataset to samples that only contain observations with common support over observables (\cite{crump2009dealing}). We use logistic regressions to restrict our dataset to a common support. More specifically, we estimate:

\begin{equation}
      \Phi(M_{ict}) = \beta_0 + \theta X_{ic}+\epsilon_{ic}
\end{equation}
where $M_{ict}$ is an indicator variable equal to one if observation $i$ in county $c$ in period $t$ is (or resides in) a majority black neighborhood and $X_{ict}$ is a vector of other controls for observation $i$ in county $c$ in period $t$.  Controls included quintiles of the credit score distributions, the fraction of IRS filings under \$25,000, between \$25,000-50,000, \$50,000-75,000, \$75,000-100,000, and over \$100,000, the average income, median credit scores, the Gini index of income inequality, 90+ days-past-due debt balances, unemployment and divorce rates, population density, median house value, and education attainment levels such as the fraction with at least a bachelors degree or with less than a high school diploma.

\subsection{Sensitivity Analysis}\label{appendixa2}

\subsubsection{Differences in Income and Credit Score}

In this section, we show how the coefficient on majority black changes as we add different controls to our neighborhood-level regression specification. 
 Column (1) of Table \ref{tab:creditworthiness} shows that majority-black neighborhoods have about 1.4 more judgments per every 100 people than non-black neighborhoods controlling only for county and year fixed effects. This coefficient implies that the baseline judgment rate in black neighborhoods is more than double that of non-majority black neighborhoods, where the average judgment rate is 1.26 judgments per 100 people. We progressively add additional control variables to our baseline specification to explore if and to what extent neighborhood-level income and credit score measures are driving the relationship between a neighborhood's racial composition and the number of judgments per 100 individuals. Column (2) of Table \ref{tab:creditworthiness} adds controls for average ZIP code level income and the fraction of IRS filings under \$25,000, between \$25,000-50,000, \$50,000-75,000, \$75,000-100,000, and over \$100,000, to the previous specification.  Column (3) in Table \ref{tab:creditworthiness} adds credit score quintiles and median credit score to the baseline specification. Column (4) adds income and credit score controls into the same specification. Column (5) adds controls for total delinquent debt balances, unemployment rate, median house value, the fraction of the population with a college education, the divorce rate, and population density. In this specification, we see that majority-black neighborhoods have a 56\% higher judgment rate than non-black neighborhoods. Column (6) uses a one-year lag of our income, credit score, and baseline controls, and the coefficient of interest changes only slightly.  

\subsubsection{Differences in Plaintiff Type}\label{appendixa22}

We explore if a specific plaintiff category drives differences in judgment rates across black and non-black zip codes.  For each zip code, we know the number of judgments awarded to the following plaintiff types:  auto, debt buyer, high-cost lender, major bank, medical, utility, and miscellaneous.  Debt buyers account for 48\% of plaintiffs in our sample.  Medical lenders, major banks, and high-cost lenders are the next largest plaintiff categories accounting for 20\%, 13\%, and 6\% of plaintiffs, respectively.  The other plaintiff categories are combined into the miscellaneous category.    

Our results are presented in Table \ref{tab:plaintiff_type}.  Each specification includes the income, credit score, and baseline controls discussed in Table \ref{tab:creditworthiness}, as well as county and year fixed effects. Column (1) repeats the main analysis and includes judgments from all plaintiff types (this is the same result presented in Column (5) of Table \ref{tab:creditworthiness}). Column (2) limits the outcome variable to only judgments obtained by debt buyers, Column (3) to major banks, Column (4) to medical companies, Column (5) to high-cost lenders, and Column (6) to any other lender.  

The coefficients estimated across each specification should not be directly compared due to differences in the baseline judgment rates in non-black neighborhoods across these different plaintiff types. For example, the judgment rate in non-black neighborhoods was 0.57 judgments per 100 people for debt buyers, 0.14 judgments per 100 people for miscellaneous lenders, and 0.08 judgments per 100 people for high-cost lenders. These baseline levels imply that majority-black neighborhoods have a 38\% higher judgment rate than non-black neighborhoods among debt buyers, 13\% higher among major banks, and a 156\% higher judgment rate among high-cost lenders. The racial gap in debt collection judgments is not prevalent among medical lenders. In our estimation, the miscellaneous lenders category consistently showed the highest racial imbalance in their lawsuits, followed closely by high-cost lenders.  

\subsubsection{Banks and Payday Lenders Neighborhood Analysis}

In this section, we explore differences in lending institutions across black and non-black neighborhoods at the zipcode-level.\footnote{The presence of payday lenders is likely the result of financial distress as opposed to the cause.  As such, we view these controls as an additional measure of the financial well-being of neighborhoods instead of a measure of credit access.} We use ArcGIS to create an index that measures the number of banks and payday lenders within five and 10-mile radii from each zip code's centroid.  We also use broadband access as a proxy for access to online credit markets and the share of credit reports unscored as a proxy for access to credit.\footnote{An unscored credit report is one in which there is not enough information on a consumer's credit report to issue a formal credit score.  This variable could serve as a measure of access to credit if unscored reports correlate with limited access to credit markets instead of a limited desire to obtain credit.}  We add these variables as controls to our main specification.  

This analysis only used data from 2008-2012, and thus our sample size is slightly smaller.  For context, we replicate the main results on this subsample of the data in Column (1) of Table \ref{tab:access_to_credit}.  Columns (2) and (3) add controls for broadband access and the share of unscored accounts in a zip code; both measures have no meaningful impact on our coefficient of interest.  Columns (4) and (5) add our controls for the number of banks and payday lenders within five and 10-mile radii.  We see that the number of banks negatively correlates with the judgment rate, while the number of payday lenders positively correlates with the number of judgments.  Adding these controls decreases the coefficient on the black majority by 14\%, though a large racial gap in the number of judgments issued across black and non-black communities remains.

\subsubsection{Alternative Credit Score and Other Additional Controls}

In Table \ref{tab:pp}, we replicate Table \ref{tab:creditworthiness} but add an alternative control for credit score calculated using a deep learning algorithm. The model consistently outperforms standard credit scoring models when predicting default rates (\cite{albanesi_vamossy}). This alternative credit score has more predictive power than the conventional credit score in predicting default. However, it does not mitigate the racial disparity we see in judgments across black and non-black communities. This finding supports that differences in credit scores, which measure a borrower's likelihood of default, are not the main factor driving the judgment gap between black and non-black communities.  

Each of our main specifications includes a static measure of housing values to control for differences in housing-related wealth across majority black and non-black neighborhoods. However, it is likely the case that neighborhoods experience differential housing market dynamics. To address concerns that our static neighborhood-level measures of housing values could miss such dynamic differences across majority black and majority non-black neighborhoods, we add Zillow's annual zip code level measure of home prices to our model. We also know that differential mortgage denial rates exist across white and black neighborhoods (\cite{bartlett2021consumer}). We, therefore, supplement our model with HMDA data documenting mortgage denial rates by zip code to see if such differences can explain the disparity in debt collection judgments across black and non-black neighborhoods. 

Table \ref{tab:access_to_housing} presents these results.  Column (1) shows our preferred results on the subsample of data for which HMDA data is available, and Column (3) shows our preferred specification on the subsample of data for which Zillow data is available. Columns (2) and (4) show how our coefficient of interest changes when HMDA and Zillow data are added to the empirical model. Adding these controls decreases the coefficient on majority-black by 30\%, though a large racial gap in the number of judgments issued across black and non-black communities remains.

\subsubsection{Alternative Dependent Variables and Weighting}

Throughout our main neighborhood-level analysis, we focus on the number of judgments per 100 individuals as our primary outcome variable of interest. The results from that analysis suggest there are more judgments per 100 people in majority-black neighborhoods than in majority non-black neighborhoods. In Column (1) of Table \ref{tab:judgment_alternative} we instead document the share of cases that resulted in a judgment. We see that majority-black neighborhoods are three percentage points more likely to have a case result in a judgment. A lower share of settled cases before trial is the primary driving force behind this difference. Because settling cases pre-trial often requires a large lump sum payment, this finding provides suggestive evidence that defendants from majority non-black neighborhoods can better mitigate negative shocks.\footnote{Since settling a case often requires a one-time lump sum payment, defendants who settle tend to have worse subsequent credit outcomes (Cheng et al. 2019). This finding suggests that a lower propensity to settle cases before a trial could help defendants from majority-black neighborhoods in the long run.}  

We further alter our outcome variable of interest by normalizing by the number of judgments per 100 people with tax filings under \$25,000 as opposed to normalizing by the total population. This specification presents an alternative way to address concerns that the risk set of individuals potentially exposed to debt collection judgments differs across black and non-black neighborhoods. The results are presented in Columns (2) of Table \ref{tab:judgment_alternative}.  Column (4) presents similar results but with the number of judgments normalized by 100 people with subprime or worse credit. While these coefficients can not be directly compared to our previous results due to differences in the sample averages of judgments per 100 people and judgments per 100 people with an increased likelihood of default, we still see positive and statistically significant coefficients on our black majority indicator variable.  

Column (6) of Table \ref{tab:judgment_alternative} presents results that winsorized our primary outcome variable at the 99th percentile to mitigate the impact of outliers. The positive and statistically significant coefficient suggests that outliers are not driving our main findings. Lastly, we explore the impact of not using population weighting. Our main neighborhood-level analysis uses population weighting since we are interested in the aggregate and representative gap. However, as shown in Table \ref{tab:summary_statistics}, black zip codes are more populated than non-black zip codes, meaning the judgment gap could be magnified by (or potentially solely driven by) the decision to population weight our regressions. As such, we also run equally weighted regressions. We consistently find coefficients that are of very similar signs, magnitudes, and statistical significance, as evidenced by Columns (3), (5), and (7) that replicate our previous three results.

\subsubsection{Neighborhood Racial Composition}\label{appendixa26}

In this section, we show that our results are robust to using the share of black residents in a neighborhood instead of a binary measure. Columns (1) and (2) of Table \ref{tab:race_proxies} present this result.  Column (1) shows a positive and statistically significant coefficient on the share of black residents within a zip code. This result suggests that a neighborhood with only black residents would have 1.5 more judgments per 100 people than a similar neighborhood with no black residents, more than double the baseline judgment rate in non-majority black neighborhoods. Column (2) shows our preferred specification from Table \ref{tab:creditworthiness}, which uses our binary measure for a black neighborhood for comparison. Tables \ref{tab:debt_composition_share_black} and \ref{tab:judgment_type_share_black} in the appendix replicate our main neighborhood-level results (Tables \ref{tab:debt_composition} and \ref{tab:judgment_type} in the main text) using the share of black residents instead of a binary measure. Results are statistically and economically similar to the results from our primary analysis.    

Since many creditors do not collect racial demographic information, the main analysis focuses on disparities across majority black and majority non-black neighborhoods. One could explore the racial composition of defendants within a neighborhood; however, since racial demographic information is often not collected by creditors, it is not available in the court data. 

We instead use a Bayesian Improved Surname Geocoding (BISG) to estimate the racial composition of the defendant pool (Elliott et al., 2009).\footnote{Consumer Finance Protection Bureau’s Office of Research (OR), the Division of Supervision, Enforcement, and Fair Lending (SEFL) rely on a Bayesian Improved Surname Geocoding (BISG) proxy method.} This method combines publicly available geography- and surname-based information into a single proxy probability for an individual's race using the Bayes updating rule. This method involves constructing a probability of assignment to race based on demographic information associated with surname and then updating this probability using the demographic characteristics of the zip code associated with place of residence. The updating is performed through a Bayesian algorithm, which yields an integrated probability that can be used to proxy for an individual’s race and ethnicity. We again classify the defendant pool as majority-black if at least 50\% of the defendants in the debt collection data were predicted to be black. Information on the racial and ethnic composition of the U.S. population by geography comes from the Summary File 1 (SF1) from the 2010 Census, which provides counts of enumerated individuals by age, race, and ethnicity for various geographic area definitions including zip code tabulation areas.\footnote{Census blocks are the highest level of disaggregation (the smallest geography).} Research has found that this approach produces proxies that correlate highly with self-reported race and national origin and is more accurate than relying only on demographic information associated with a borrower’s last name or place of residence alone (CFBP Report, 2014).   

Vectors of six racial/ethnic probabilities for each listed surname (corrected for suppression and low-frequency surnames) are used as the first input into the BISG algorithm. This information calculates a prior probability of an individual's race/ethnicity. The algorithm updates these prior probabilities with geocoded ZCTA proportions for these groups from the 2010 Census SF1 files to generate posterior probabilities. Let J equal the number of names on the enhanced surname list plus one to account for names, not on the list and let K equal the number of ZCTA in the 2010 census with any population. We define the prior probability of a person’s race based on surname so that for a person with surname j = 1, ..., J on the list, the prior probability for the race, i = 1,...,6, is p(i|j) = proportion of all people with surname j who report being of the race i in the enhanced surname file (the probability of a selected race given surname). This probability is updated based on ZCTA residence. For ZCTA k = 1,...,K, r(k|i) = proportion of all people in redistributed SF1 file who self report being race i who reside in ZCTA k (the probability of a selected ZCTA of residence given race/ethnicity).  Let $u(i, j, k) = p(i|j) * r(k|i)$.  According to Bayes' Theorem and the assumption that the probability of residing in a given ZCTA given a person's race does not vary by surname, the updated (posterior) probability of being of race/ethnicity i given surname j and ZCTA of residence k can be calculated as follows:

\begin{equation}
 q(i|j,k)=\frac{u(i,j,k)}{u(1,j,k)+u(2,j,k)+u(3,j,k)+u(4,j,k)+u(5,j,k)+u(6,j,k)}  
\end{equation}

\noindent
Note that all parameters needed for BISG posterior probabilities are derived only from Census 2010 data and none from administrative sources.  
Columns (3)-(4) of Table A9 present the results. Once again, the results are positive and statistically significant.  We also replicate Table \ref{tab:debt_composition} using our BISG estimate of the racial composition of the defendant pool in Table \ref{tab:debt_composition_bisg}.  We find results that are similar in size and statistical significance.

\subsubsection{Selection on Unobservables}

We also investigated the impact of selection on unobservables on coefficient stability (\cite{oster_unobservable}). In particular, we used Column (5) of Table \ref{tab:creditworthiness} as our benchmark and found that given a selection on unobservables that is half the size of the selection on observables, our coefficient on majority-black is reduced to 0.53 with a 95\% confidence interval ranging from [0.26 to 0.81].\footnote{We bootstrapped our treatment coefficient estimates 100 times and assumed a maximum $R^2$ value of 0.9.}  This suggests that majority-black neighborhoods experience approximately 40\% more judgments than non-majority black neighborhoods.\footnote{We also examined the proportion of selection of unobservables to observables that would explain away our treatment effect. We found that a ratio of 1.91 with a 95\% confidence interval ranging from [0.89, 2.92] is sufficient to explain away our findings.}  This finding suggests that a racial gap is unlikely to be zero, even after controlling for unobservable characteristics.

\subsubsection{Additional Judgment Data Sources}

Lastly, we replicate our main results with zip code data from New Jersey and Cook County, Illinois. The results are presented in Table \ref{tab:summary_statistics_il_nj} and \ref{tab:debt_composition_nj_il}, which replicate Tables \ref{tab:summary_statistics} and \ref{tab:debt_composition} using data from New Jersey and Cook County, Illinois.\footnote{We cannot replicate Table \ref{tab:judgment_type} with this data as this information is not available for this sample.}  Table \ref{tab:summary_statistics_il_nj} shows that judgments per 100 people are larger in majority-black neighborhoods compared to majority non-black neighborhoods, while median income and median credit score tend to be lower. Table \ref{tab:debt_composition_nj_il} shows that credit scores, income levels, and debt characteristics are not driving the racial disparity in debt collection judgments. These results suggest that judgments are 18\% higher in black neighborhoods compared to non-black neighborhoods.  

We also relax the common support assumption and replicate our main results on the entire Missouri sample. These results are presented in tables \ref{tab:summary_statistics_off_common} through \ref{tab:judgment_type_off_common}.  Again, we find very similar results to our main specification, which uses only the common support sample, suggesting that omitted variables correlated with observable neighborhood characteristics are not biasing our results in any particular direction.

\subsubsection{Other Demographics}

In Table \ref{tab:race_other}, we replicate Table \ref{tab:creditworthiness} with additional controls for the share of the Hispanic and Asian population within each zip code. We do not observe any consistent judgment gap for Hispanic across our various specifications. However, we find that the share of the Asian population is negatively related to judgments. The share of black residents remains positive and statistically significant in each of the specifications.

\subsubsection{Non-linearities \& Higher Order Interactions}

Lastly, we investigate if causal machine learning techniques that allow for high-order interactions of observable characteristics can mitigate the disparity. This relies on the assumption that more information can be extracted from the interaction of multiple variables. More specifically, we adopt Double Machine Learning by \cite{chernozhukov2018double}, and implement Gradient Boosted Trees (GBT) and Random Forests in the first stages, with linear regression and robust standard errors in the second.\footnote{For more information about the GBT model, see \cite{friedman}.} Gradient Boosted Trees (GBT) is an ensemble learning approach that mitigates the tendency of tree-based models' to overfit training data. This is accomplished by recursively combining the forecasts of many over-simplified trees. The theory behind boosting proposes that a collection of weak learners as an ensemble create a single strong learner with improved stability over a single complex tree. 

At each step m, $1 \le m \le M$, of gradient boosting, an estimator, $h_m$, is computed on the residuals from the previous models' predictions. A critical part of the gradient boosting method is regularization by shrinkage as proposed by \cite{friedman}. This consists in modifying the update rule as follows:
\begin{equation}
    F_m(x) = F_{m-1}(x) + \nu \gamma_m h_m(x),
\end{equation}
where $h_m(x)$ represents a weak learner of fixed depth, $\gamma_m$ is the step length and $\nu$ is the learning rate or shrinkage factor.

The estimation procedure begins with fitting a shallow tree (e.g., with depth L = 1). Using the prediction residuals from the first tree, fit a second tree with the same shallow depth L. Weight the predictions of the second tree by $\nu \in (0,1)$ to prevent the model from overfitting the residuals, and then aggregate the forecasts of these two trees. At each step k, fit a shallow tree to the residuals from the model with k-1 trees, and add its prediction to the forecast of the ensemble with a shrinkage weight of $\nu$. Do this until a total of K trees is reached in the ensemble. For our GBT model, we split the data into three chunks: training set (60\%), holdout set (20\%), and testing set (20\%). We relied on XGBoost for the implementation of our GBT model (\cite{xgboost}).  We find that Double Machine Learning is unable to explain away the disparity. In particular, the estimates obtained in Table \ref{tab:dml} are similar to the ones obtained in Table \ref{tab:creditworthiness}.

\clearpage

\subsection*{Appendix Figures}\label{sec:fig}
\addcontentsline{toc}{section}{Figures}

\begin{figure}[!h]
\footnotesize
\begin{center}
\begin{subfigure}[b]{0.7\textwidth}
\includegraphics[width=\textwidth]{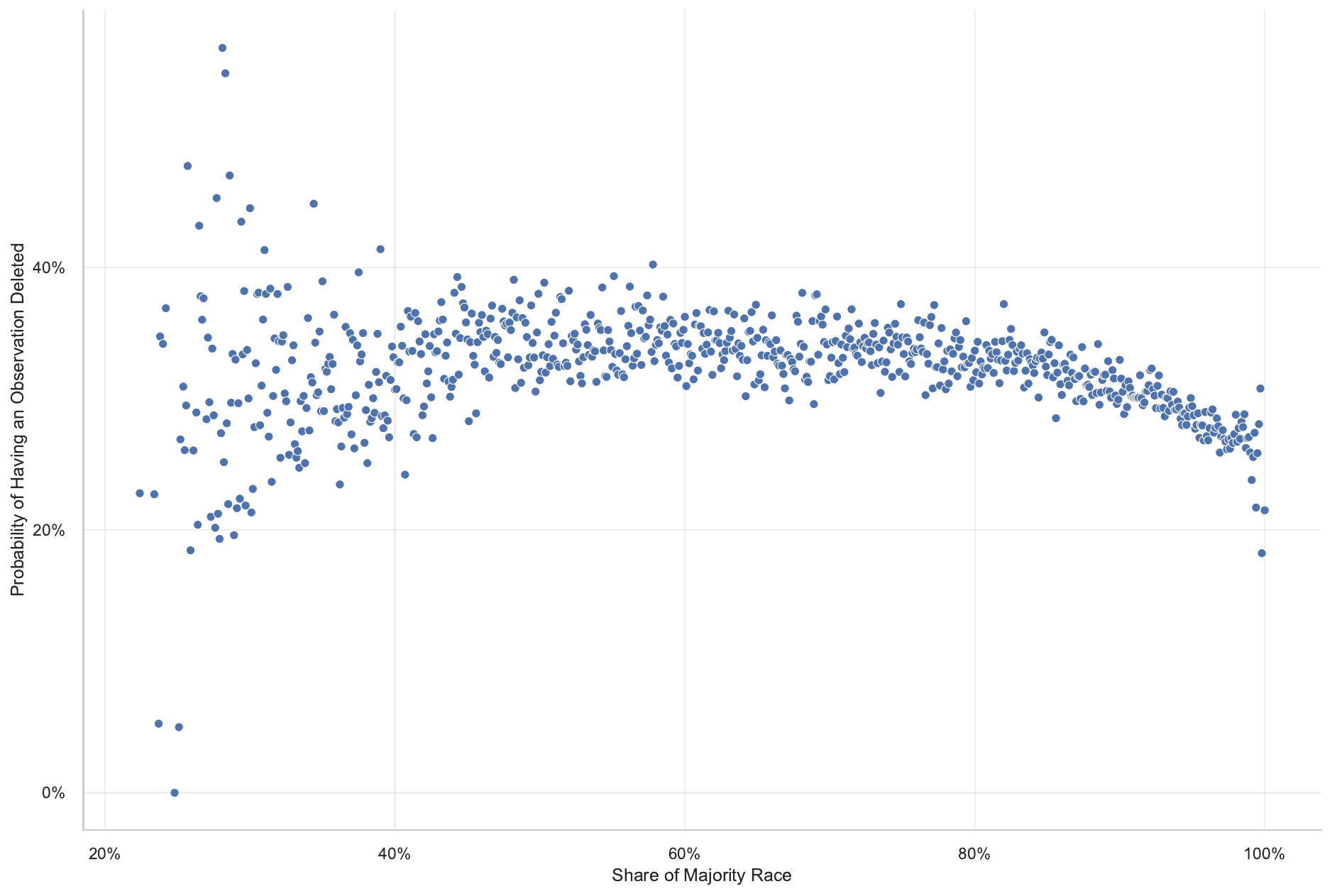}
\caption{}
\end{subfigure}
\begin{subfigure}[b]{0.7\textwidth}
\includegraphics[width=\textwidth]{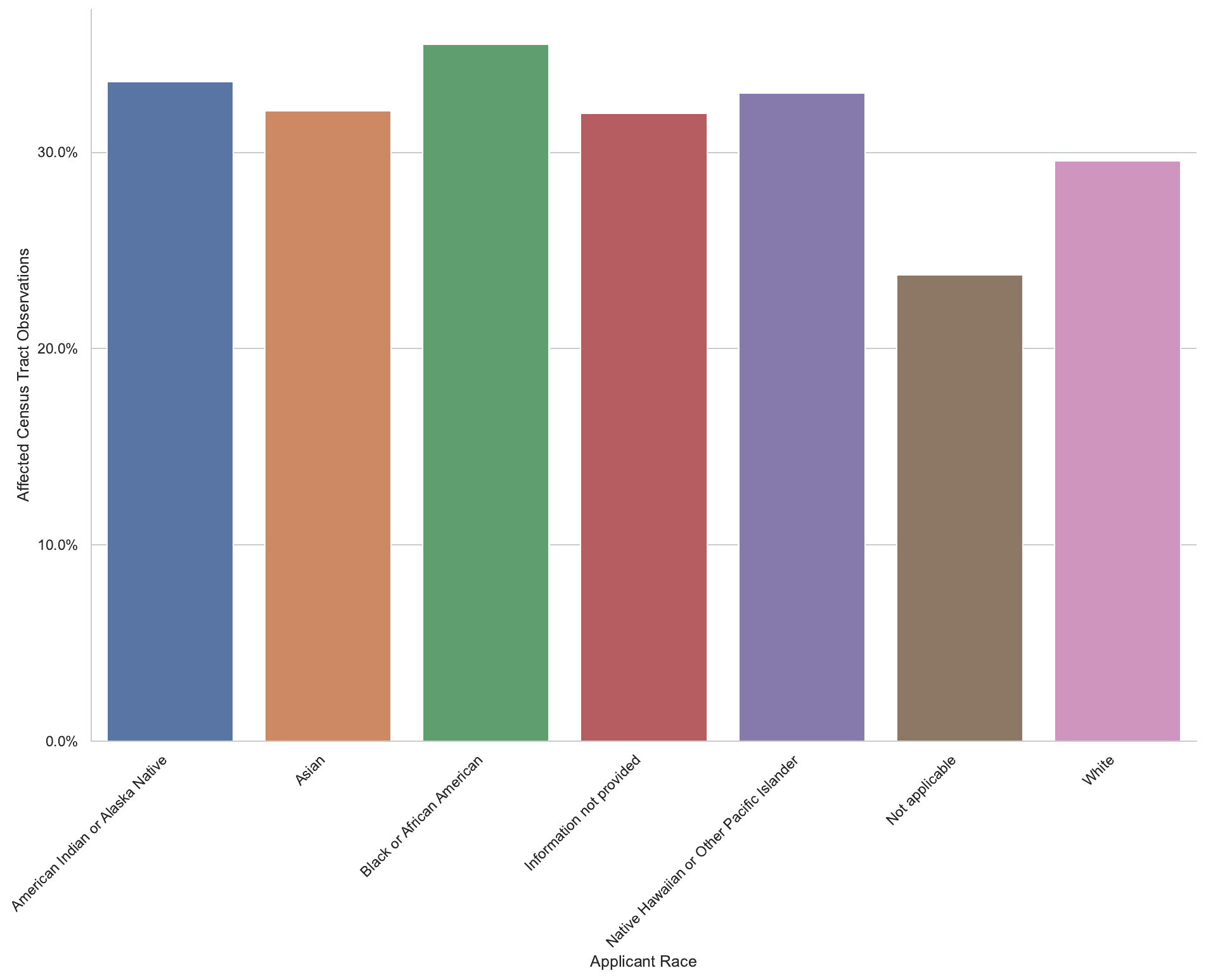}
\caption{}
\end{subfigure}
\caption{Duplicate Observations by (a) Share of Majority Race and (b) Race}\label{fig:deletion_by_majority_share}
\end{center}
\begin{flushleft}
\scriptsize{Notes: This figure illustrates the probability of having an observation deleted during the cleaning and merging of credit report data with HMDA data as a function of (a) the share of the majority race in a given Census Tract and (b) race. Deletions are more common in ZIP codes with more diverse populations.}
\end{flushleft}
\end{figure}

\clearpage 

\subsection*{Appendix Tables}
\input{Tables/Tables_neighborhood/sample_restrictions}

\input{Tables/Tables_neighborhood/results_judgments_neighborhood}
\input{Tables/Tables_neighborhood/results_plaintiff_type_neighborhood}
\input{Tables/Tables_neighborhood/results_credit_markets_neighborhood}
\input{Tables/Tables_neighborhood/results_alternative_credit_risk_neighborhood}
\input{Tables/Tables_neighborhood/results_housing_market_neighborhood}

\input{Tables/Tables_neighborhood/results_alternative_dependent_neighborhood}
\input{Tables/Tables_neighborhood/results_race_proxies_neighborhood}

\input{Tables/Tables_share/results_debt_portfolio_neighborhood_share_black}
\input{Tables/Tables_share/results_judgment_type_neighborhood_share_black}

\input{Tables/Tables_neighborhood/results_debt_portfolio_neighborhood_bisg}

\input{Tables/Tables_neighborhood/summary_tab1_other_areas}
\input{Tables/Tables_neighborhood/results_debt_portfolio_neighborhood_other_areas}

\input{Tables/Tables_noncommon/summary_tab1_off_common_support}
\input{Tables/Tables_noncommon/results_debt_portfolio_neighborhood_off_common_support}
\input{Tables/Tables_noncommon/results_judgment_type_neighborhood_off_common_support}

\input{Tables/Tables_neighborhood/results_other_races_neighborhood}
\input{Tables/Tables_neighborhood/dml}
\clearpage

%% file: Tables/Tables_neighborhood/sample_restrictions.tex
\begin{table}[htbp]\centering
\begin{threeparttable}
\footnotesize
\caption{Itemized Sample Restrictions: Missouri Data}\label{tab:sample_restrictions}
\begin{tabular}{lcc} \hline \hline
 &  ZIP Codes & \% Black  \\ \hline 
ACS & 1,021 &  4.8\%  
 \\ 
Demographic \& Judgment Information Available  & 1,015 &  4.8\%
 \\ 
Credit Report Information Available  & 902 &  5.5\%
 \\ 
Common Support Sample  & 840 &  5.6\%
 \\ 
 \hline \hline 
\end{tabular}
\begin{tablenotes}
\scriptsize
\item Notes:  This table documents the numbers of zip codes and the average percentage of the population that is black across different samples of data from Missouri.  
\end{tablenotes}
\end{threeparttable}
\end{table}

%% file: Tables/Tables_neighborhood/results_judgments_neighborhood.tex
\begin{table}[htbp] \centering
\begin{threeparttable}
\footnotesize
\def\sym#1{\ifmmode^{#1}\else\(^{#1}\)\fi}
\caption{Judgments, Income, and Credit Scores \label{tab:creditworthiness}}
\begin{tabular}{l*{6}{c}}
\hline\hline
                    &\multicolumn{1}{c}{(1)}&\multicolumn{1}{c}{(2)}&\multicolumn{1}{c}{(3)}&\multicolumn{1}{c}{(4)}&\multicolumn{1}{c}{(5)}&\multicolumn{1}{c}{(6)}\\
                       &\multicolumn{6}{c}{Judgment Rate}\\
\hline
& & & & & & \\[\dimexpr-\normalbaselineskip+2pt]
Black Majority: ZIP &      1.4356\sym{***}&      0.8629\sym{***}&      1.0786\sym{***}&      0.7939\sym{***}&      0.7084\sym{***}&      0.8384\sym{***}\\
                    &    (0.1873)         &    (0.1917)         &    (0.1658)         &    (0.1807)         &    (0.1662)         &    (0.0356)         \\
\hline
& & & & & & \\[\dimexpr-\normalbaselineskip+2pt]

Baseline Controls   &                     &                     &                     &                     &           X         &                     \\
Income      &                     &           X         &                     &           X         &           X         &                     \\
Credit Score    &                     &                     &           X         &           X         &           X         &                     \\
Lagged Baseline Controls&                     &                     &                     &                     &                     &           X         \\
& & & & & & \\[\dimexpr-\normalbaselineskip+2pt] \hline 
Wild Cluster Bootstrap p-value&           0         &           0         &           0         &       .0005         &       .0009         &           0         \\
Observations        &   7365      &   7365      &   7019      &   7019      &   7019      &   6105      \\
$R^2$               &      0.5961         &      0.6617         &      0.6279         &      0.6684         &      0.6851         &      0.6770         \\
\hline\hline
\end{tabular}
\begin{tablenotes}
\scriptsize
\item Notes:  This table presents results of a regression of the judgment rate on neighborhood racial composition and other neighborhood-level characteristics.  The sample includes Missouri zip codes in the common support sample.  The dependent variable is judgments per 100 individuals.  Income controls include average zip code-level income and the fraction of IRS filings under \$25,000, between \$25,000-50,000, \$50,000-75,000, \$75,000-100,000, and over \$100,000; credit controls include credit score quintiles and median credit score; baseline controls include total delinquent debt balances, unemployment rate, median house value, the fraction of the population with a college education, the divorce rate, and population density. All regressions are weighted by population. Robust standard errors clustered at the county-year level are in parentheses. Effective number of clusters: 36.0. Wild Cluster Bootstrap p-values are reported for our main parameter of interest. \sym{*} \(p<0.10\), \sym{**} \(p<0.05\), \sym{***} \(p<0.01\). 
\end{tablenotes}
\end{threeparttable}
\end{table}

%% file: Tables/Tables_neighborhood/results_plaintiff_type_neighborhood.tex
\begin{table}[htbp] \centering
\begin{threeparttable}
\footnotesize
\def\sym#1{\ifmmode^{#1}\else\(^{#1}\)\fi}
\caption{Judgments and Plaintiff Type\label{tab:plaintiff_type}}
\begin{tabular}{l*{6}{c}}
\hline\hline
                    &\multicolumn{1}{c}{(1)}&\multicolumn{1}{c}{(2)}&\multicolumn{1}{c}{(3)}&\multicolumn{1}{c}{(4)}&\multicolumn{1}{c}{(5)}&\multicolumn{1}{c}{(6)}\\
                    &\multicolumn{1}{c}{Judgments}&\multicolumn{1}{c}{Debt Buyer}&\multicolumn{1}{c}{Major Bank}&\multicolumn{1}{c}{Medical}&\multicolumn{1}{c}{High-Cost }&\multicolumn{1}{c}{Misc.}\\
\hline
& & & & & & \\[\dimexpr-\normalbaselineskip+2pt]
Black Majority: ZIP &       0.708\sym{***}&       0.217\sym{***}&       0.058\sym{**} &       0.045         &       0.119\sym{***}&       0.321\sym{***}\\
                    &     (0.166)         &     (0.051)         &     (0.029)         &     (0.055)         &     (0.024)         &     (0.082)         \\
\hline
& & & & & & \\[\dimexpr-\normalbaselineskip+2pt]
Mean                &       1.253         & .568       &        .453         &        .297         &        .076         &        .139         \\
Effect Size         &56.5         &        38.2         &        12.8         &        15.1         &       156.3         &         231         \\ \hline
& & & & & & \\[\dimexpr-\normalbaselineskip+2pt]
Wild Cluster Bootstrap p-value&       .0009         &       .0006         &       .0652         &       .5694         &           0         &       .0001         \\
Observations        &    7019        &    7019        &    7019        &    7019        &    7019        &    7019        \\
$R^2$               &       0.685         &       0.699         &       0.725         &       0.613         &       0.602         &       0.533         \\
\hline\hline
\end{tabular}
\begin{tablenotes}
\scriptsize
\item Notes:  This table presents results of a regression of the judgment rate for specific types of debt collectors on neighborhood racial composition and other neighborhood-level characteristics.  The sample includes Missouri zip codes in the common support sample.  The dependent variable in Column (1) is judgments per 100 individuals.  The dependent variables in Columns (2)-(6) are judgements by debt buyers per 100 individuals, judgments by major banks per 100 individuals, judgments by medical lenders per 100 individuals, judgments by high-cost lenders per 100 individuals, and judgments by other miscellaneous lenders per 100 individuals. Baseline controls (average zip code-level income, the fraction of IRS filings under \$25,000, between \$25,000-50,000, \$50,000-75,000, \$75,000-100,000, and over \$100,000, credit score quintiles, median credit score, total delinquent debt balances, unemployment rate, median house value, the fraction of the population with a college education, the divorce rate, and population density), county fixed effects, and year fixed effects are included in all specifications. All regressions are weighted by population. Robust standard errors clustered at the county-year level are in parentheses. Effective number of clusters: 36.0. Wild Cluster Bootstrap p-values are reported for our main parameter of interest. \sym{*} \(p<0.10\), \sym{**} \(p<0.05\), \sym{***} \(p<0.01\). 
\end{tablenotes}
\end{threeparttable}
\end{table}

%% file: Tables/Tables_neighborhood/results_credit_markets_neighborhood.tex
\begin{table}[htbp] \centering
\begin{threeparttable}
\footnotesize
\def\sym#1{\ifmmode^{#1}\else\(^{#1}\)\fi}
\caption{Judgments and Access to Credit Markets\label{tab:access_to_credit}}
\begin{tabular}{l*{5}{c}}
\hline\hline
                    &\multicolumn{1}{c}{(1)}&\multicolumn{1}{c}{(2)}&\multicolumn{1}{c}{(3)}&\multicolumn{1}{c}{(4)}&\multicolumn{1}{c}{(5)}\\
                       &\multicolumn{5}{c}{Judgment Rate}\\
\hline
& & & & & \\[\dimexpr-\normalbaselineskip+2pt]
Black Majority: ZIP &      1.0910\sym{***}&      1.0897\sym{***}&      1.0914\sym{***}&      0.9375\sym{***}&      0.9417\sym{***}\\
                    &    (0.1398)         &    (0.1394)         &    (0.1399)         &    (0.1099)         &    (0.1135)         \\
& & & & & \\[\dimexpr-\normalbaselineskip+2pt]
Broadband           &                     &      0.0565\sym{**} &                     &                     &                     \\
                    &                     &    (0.0222)         &                     &                     &                     \\
& & & & & \\[\dimexpr-\normalbaselineskip+2pt]
Unscored            &                     &                     &      0.1838         &                     &                     \\
                    &                     &                     &    (0.2691)         &                     &                     \\
& & & & & \\[\dimexpr-\normalbaselineskip+2pt]
Banks (5 miles)     &                     &                     &                     &     -0.0056\sym{***}&                     \\
                    &                     &                     &                     &    (0.0011)         &                     \\
& & & & & \\[\dimexpr-\normalbaselineskip+2pt]
Payday Lenders (5 miles)&                     &                     &                     &      0.0177\sym{***}&                     \\
                    &                     &                     &                     &    (0.0028)         &                     \\
& & & & & \\[\dimexpr-\normalbaselineskip+2pt]
Banks (10 miles)    &                     &                     &                     &                     &     -0.0025\sym{***}\\
                    &                     &                     &                     &                     &    (0.0003)         \\
& & & & & \\[\dimexpr-\normalbaselineskip+2pt]
Payday Lenders (10 miles)&                     &                     &                     &                     &      0.0098\sym{***}\\
                    &                     &                     &                     &                     &    (0.0013)         \\
\hline
& & & & & \\[\dimexpr-\normalbaselineskip+2pt]
Wild Cluster Bootstrap p-value&           0         &           0         &           0         &           0         &           0         \\
\hline 
& & & & & \\[\dimexpr-\normalbaselineskip+2pt]
Observations        &   4183      &   4183      &   4183      &   4183      &   4183      \\
$R^2$               &      0.8181         &      0.8186         &      0.8181         &      0.8279         &      0.8295         \\
\hline\hline
\end{tabular}
\begin{tablenotes}
\scriptsize
\item Notes:  This table presents results of a regression of the judgment rate on neighborhood racial composition and other neighborhood-level characteristics.  The sample includes Missouri zip codes in the common support sample.  The dependent variable is judgments per 100 individuals.  Baseline controls (average zip code-level income, the fraction of IRS filings under \$25,000, between \$25,000-50,000, \$50,000-75,000, \$75,000-100,000, and over \$100,000, credit score quintiles, median credit score, total delinquent debt balances, unemployment rate, median house value, the fraction of the population with a college education, the divorce rate, and population density), county fixed effects, and year fixed effects are included in all specifications. All regressions are weighted by population. Robust standard errors clustered at the county-year level are in parentheses. Effective number of clusters: 36.0. Wild Cluster Bootstrap p-values are reported for our main parameter of interest. \sym{*} \(p<0.10\), \sym{**} \(p<0.05\), \sym{***} \(p<0.01\). 
\end{tablenotes}
\end{threeparttable}
\end{table}

%% file: Tables/Tables_neighborhood/results_alternative_credit_risk_neighborhood.tex
\begin{table}[htbp] \centering
\begin{threeparttable}
\footnotesize 
\def\sym#1{\ifmmode^{#1}\else\(^{#1}\)\fi}
\caption{Judgments and an Alternative Credit Score\label{tab:pp}}
\begin{tabular}{l*{5}{c}}
\hline\hline
                    &\multicolumn{1}{c}{(1)}&\multicolumn{1}{c}{(2)}&\multicolumn{1}{c}{(3)}&\multicolumn{1}{c}{(4)}&\multicolumn{1}{c}{(5)}\\
                       &\multicolumn{5}{c}{Judgment Rate}\\
\hline
& & & & &  \\[\dimexpr-\normalbaselineskip+2pt]
Black Majority: ZIP &      1.6061\sym{***}&      1.2321\sym{***}&      0.9537\sym{***}&      0.8287\sym{***}&      1.0213\sym{***}\\
                    &    (0.1885)         &    (0.1717)         &    (0.1961)         &    (0.1764)         &    (0.0360)         \\
\hline
& & & & &  \\[\dimexpr-\normalbaselineskip+2pt]
Baseline Controls   &                     &                     &                     &           X         &                     \\
Income              &                     &                     &           X         &           X         &                     \\
Default Probability &                     &           X         &           X         &           X         &           X         \\
Lagged Baseline Controls&                     &                     &                     &                     &           X         \\
\hline
& & & & &  \\[\dimexpr-\normalbaselineskip+2pt]
Wild Cluster Bootstrap p-value&           0         &           0         &       .0001         &       .0007         &           0         \\
Observations        &   5909      &   5909      &   5909      &   5625      &   4738      \\
$R^2$               &      0.5977         &      0.6319         &      0.6813         &      0.7001         &      0.7291         \\
\hline\hline
\end{tabular}
\begin{tablenotes}
\scriptsize
\item Notes:  This table presents results of a regression of the judgment rate on neighborhood racial composition and other neighborhood-level characteristics.  The sample includes Missouri zip codes in the common support sample.  The dependent variable is judgments per 100 individuals.  Baseline controls (average zip code-level income, the fraction of IRS filings under \$25,000, between \$25,000-50,000, \$50,000-75,000, \$75,000-100,000, and over \$100,000, total delinquent debt balances, unemployment rate, median house value, the fraction of the population with a college education, the divorce rate, and population density), county fixed effects, and year fixed effects are included in all specifications. Default probabilities are used as control variables in place of credit scores. Column (5) controls for lagged default probability instead of default probability. Default probability data is available starting in 2006. All regressions are weighted by population. Robust standard errors clustered at the county-year level are in parentheses. Effective number of clusters: 36.0. Wild Cluster Bootstrap p-values are reported for our main parameter of interest. \sym{*} \(p<0.10\), \sym{**} \(p<0.05\), \sym{***} \(p<0.01\). 
\end{tablenotes}
\end{threeparttable}
\end{table}

%% file: Tables/Tables_neighborhood/results_housing_market_neighborhood.tex
\begin{table}[htbp] \centering
\begin{threeparttable}
\footnotesize
\def\sym#1{\ifmmode^{#1}\else\(^{#1}\)\fi}
\caption{Judgments and Local Housing Markets\label{tab:access_to_housing}}
\begin{tabular}{l*{4}{c}}
\hline\hline
                    &\multicolumn{1}{c}{(1)}&\multicolumn{1}{c}{(2)}&\multicolumn{1}{c}{(3)}&\multicolumn{1}{c}{(4)}\\
                       &\multicolumn{4}{c}{Judgment Rate}\\
\hline
& & & & \\[\dimexpr-\normalbaselineskip+2pt]
Black Majority: ZIP &      0.9451\sym{***}&      0.8416\sym{***}&      0.7136\sym{***}&      0.6891\sym{***}\\
                    &    (0.1581)         &    (0.1453)         &    (0.1610)         &    (0.1584)         \\
& & & & \\[\dimexpr-\normalbaselineskip+2pt]
HMDA: Share Rejected&                     &      1.0981\sym{***}&                     &                     \\
                    &                     &    (0.3638)         &                     &                     \\
& & & & \\[\dimexpr-\normalbaselineskip+2pt]
Zillow: Single Family House&                     &                     &                     &     -0.0060\sym{***}\\
                    &                     &                     &                     &    (0.0013)         \\
\hline
& & & & \\[\dimexpr-\normalbaselineskip+2pt]
HMDA Available      &           X         &           X         &                     &                     \\
Zillow Available    &                     &                     &           X         &           X         \\ \hline
& & & & \\[\dimexpr-\normalbaselineskip+2pt]
Wild Cluster Bootstrap p-value&           0         &           0         &       .0006         &       .0005         \\
Observations        &   3846         &   3846         &   5231         &   5231         \\
$R^2$               &      0.7593         &      0.7628         &      0.6767         &      0.6861         \\
\hline\hline
\end{tabular}
\begin{tablenotes}
\scriptsize
\item Notes:  This table presents results of a regression of the judgment rate on neighborhood racial composition and other neighborhood-level characteristics, including information on local housing markets.  The sample includes Missouri zip codes in the common support sample.  The dependent variable is judgments per 100 individuals.  Baseline controls (average zip code-level income, the fraction of IRS filings under \$25,000, between \$25,000-50,000, \$50,000-75,000, \$75,000-100,000, and over \$100,000, credit score quintiles, median credit score, total delinquent debt balances, unemployment rate, median house value, the fraction of the population with a college education, the divorce rate, and population density), county fixed effects, and year fixed effects are included in all specifications.  All regressions are weighted by population. Robust standard errors clustered at the county-year level are in parentheses. Effective number of clusters: 36.0. Wild Cluster Bootstrap p-values are reported for our main parameter of interest. \sym{*} \(p<0.10\), \sym{**} \(p<0.05\), \sym{***} \(p<0.01\). 
\end{tablenotes}
\end{threeparttable}
\end{table}

%% file: Tables/Tables_neighborhood/results_alternative_dependent_neighborhood.tex
\begin{sidewaystable}[htbp] \centering
\begin{threeparttable}
\footnotesize
\def\sym#1{\ifmmode^{#1}\else\(^{#1}\)\fi}
\caption{Judgments and Alternative Dependent Variables\label{tab:judgment_alternative}}
\begin{tabular}{l*{7}{c}}
\hline\hline
                    &\multicolumn{1}{c}{(1)}&\multicolumn{1}{c}{(2)}&\multicolumn{1}{c}{(3)}&\multicolumn{1}{c}{(4)}&\multicolumn{1}{c}{(5)}&\multicolumn{1}{c}{(6)}&\multicolumn{1}{c}{(7)}\\
&\multicolumn{1}{c}{Judgments/Cases}&\multicolumn{2}{c}{Judgment: Low Income}&\multicolumn{2}{c}{Judgment: Low Credit }&\multicolumn{2}{c}{Judgment: Winsorized} \\
\hline
& & & & & &  \\[\dimexpr-\normalbaselineskip+2pt]
Black Majority: ZIP &    0.028\sym{***} &       2.843\sym{***}&       2.649\sym{***}&       0.503\sym{***}&       0.492\sym{***} &      0.708\sym{***}&       0.666\sym{***}\\
              &     (0.010)                   &     (0.704)         &     (0.570)         &     (0.122)         &     (0.115)     &     (0.165)         &     (0.138)         \\
\hline
& & & & & &  \\[\dimexpr-\normalbaselineskip+2pt]
Population Weighted &                   &             X       &                  &                X     &                 &          X           &                     \\
\hline
& & & & & &  \\[\dimexpr-\normalbaselineskip+2pt]
Wild Cluster Boostrap p-value   &       .0111        &       .0011         &           0         &        .003         &       .0005          &       .0008         &       .0002       \\
Observations        &    7019         &    7019         &    7019              &    6999         &    6999      &    7019       &    7019         \\
$R^2$         &       0.173                &       0.660         &       0.548         &       0.744         &       0.576         &       0.686         &       0.512        \\
\hline\hline
\end{tabular}
\begin{tablenotes}
\scriptsize
\item  Notes:  This table presents results of a regression of the judgment rate on neighborhood racial composition and other neighborhood-level characteristics, where the judgment rate is normalized in various ways.  The sample includes Missouri zip codes in the common support sample.  For Column (1) our dependent variable is the fraction of cases that result in a judgment, for Columns (2-3), our outcome variable is judgments per 100 individuals with tax filings under \$25,000 USD, for Columns (4-5) our outcome variable is judgments per 100 individuals with subprime credit, and for Columns (6-7) we winsorize our main outcome variable, judgments per 100 individuals, at the 99th percentile. Baseline controls (average zip code-level income, the fraction of IRS filings under \$25,000, between \$25,000-50,000, \$50,000-75,000, \$75,000-100,000, and over \$100,000, credit score quintiles, median credit score, total delinquent debt balances, unemployment rate, median house value, the fraction of the population with a college education, the divorce rate, and population density), county fixed effects, year fixed effects, and attorney representation are included in all specifications.  Only columns (2), (4), and (6) are weighted by population. Robust standard errors clustered at the county-year level are in parentheses. Effective number of clusters: 36.0. Wild Cluster Bootstrap p-values are reported for our main parameter of interest. \sym{*} \(p<0.10\), \sym{**} \(p<0.05\), \sym{***} \(p<0.01\). 
\end{tablenotes}
\end{threeparttable}
\end{sidewaystable}

%% file: Tables/Tables_neighborhood/results_race_proxies_neighborhood.tex
\begin{table}[htbp] \centering
\begin{threeparttable}
\footnotesize
\def\sym#1{\ifmmode^{#1}\else\(^{#1}\)\fi}
\caption{Judgments and an Alternative Proxy for Race\label{tab:race_proxies}}
\begin{tabular}{l*{4}{c}}
\hline\hline
                    &\multicolumn{1}{c}{(1)}&\multicolumn{1}{c}{(2)}&\multicolumn{1}{c}{(3)}&\multicolumn{1}{c}{(4)}\\
                       &\multicolumn{4}{c}{Judgment Rate}\\
\hline
& & &  & \\[\dimexpr-\normalbaselineskip+2pt]
Share Black: ZIP    &      1.4731\sym{***}&                     &                     &                     \\
                    &    (0.2865)         &                     &                     &                     \\
& & &  & \\[\dimexpr-\normalbaselineskip+2pt]
Black Majority: ZIP &                     &      0.7084\sym{***}&                     &                     \\
                    &                     &    (0.1662)         &                     &                     \\
& & &  & \\[\dimexpr-\normalbaselineskip+2pt]
Share Black: BISG   &                     &                     &      1.3776\sym{***}&                     \\
                    &                     &                     &    (0.2563)         &                     \\
& & &  & \\[\dimexpr-\normalbaselineskip+2pt]
Black Majority: BISG&                     &                     &                     &      0.6283\sym{***}\\
                    &                     &                     &                     &    (0.1347)         \\
\hline
& & &  & \\[\dimexpr-\normalbaselineskip+2pt]
Wild Cluster Bootstrap p-value&       .0001         &       .0009         &           0         &       .0005         \\
Observations        &   7019         &   7019         &   7015         &   7015         \\
$R^2$               &      0.6965         &      0.6851         &      0.6981         &      0.6860         \\
\hline\hline
\end{tabular}
\begin{tablenotes}
\scriptsize
\item Notes:  This table presents results of a regression of the judgment rate on neighborhood racial composition (where racial composition is measured in various ways) and other neighborhood-level characteristics.  The sample includes Missouri zip codes in the common support sample.  The dependent variable is judgments per 100 individuals.  Baseline controls (average zip code-level income, the fraction of IRS filings under \$25,000, between \$25,000-50,000, \$50,000-75,000, \$75,000-100,000, and over \$100,000, credit score quintiles, median credit score, total delinquent debt balances, unemployment rate, median house value, the fraction of the population with a college education, the divorce rate, and population density), county fixed effects, and year fixed effects are included in all specifications.  All regressions are weighted by population. Robust standard errors clustered at the county-year level are in parentheses. Effective number of clusters: 36.0. Wild Cluster Bootstrap p-values are reported for our main parameter of interest. \sym{*} \(p<0.10\), \sym{**} \(p<0.05\), \sym{***} \(p<0.01\). 
\end{tablenotes}
\end{threeparttable}
\end{table}

%% file: Tables/Tables_share/results_debt_portfolio_neighborhood_share_black.tex
\begin{table}[htbp] \centering
\begin{threeparttable}
\footnotesize
\def\sym#1{\ifmmode^{#1}\else\(^{#1}\)\fi}
\caption{Judgments and Debt Portfolios: Share Black\label{tab:debt_composition_share_black}}
\begin{tabular}{l*{5}{c}}
\hline\hline
                    &\multicolumn{1}{c}{(1)}&\multicolumn{1}{c}{(2)}&\multicolumn{1}{c}{(3)}&\multicolumn{1}{c}{(4)}&\multicolumn{1}{c}{(5)}\\
                       &\multicolumn{5}{c}{Judgment Rate}\\
\hline
& & & & & \\[\dimexpr-\normalbaselineskip+2pt]
Share Black: ZIP    &      1.4651\sym{***}&      1.4758\sym{***}&      1.4727\sym{***}&      1.5054\sym{***}&      1.4915\sym{***}\\
                    &    (0.2789)         &    (0.2851)         &    (0.2850)         &    (0.2750)         &    (0.2655)         \\
\hline
& & & & & \\[\dimexpr-\normalbaselineskip+2pt]
Debt Levels         &           X         &                     &                     &                     &           X         \\
Monthly Payment and Utilization&                     &           X         &                     &                     &           X         \\
Debt Composition    &                     &                     &           X         &                     &           X         \\
Delinquency/Bankruptcy/Collections&                     &                     &                     &           X         &           X         \\
\hline
& & & & & \\[\dimexpr-\normalbaselineskip+2pt]
Wild Cluster Bootstrap p-value&           0         &       .0001         &           0         &           0         &           0         \\
Observations        &   7019        &   7019        &   7019        &   7019        &   7019        \\
$R^2$               &      0.6995         &      0.6973         &      0.6968         &      0.6995         &      0.7045         \\
\hline\hline
\end{tabular}
\begin{tablenotes}
\scriptsize
\item Notes:  This table presents results of a regression of the judgment rate on neighborhood racial composition (as measured by the share of black residents) and other neighborhood-level characteristics.  The sample includes Missouri zip codes in the common support sample.  The dependent variable is judgments per 100 individuals.  Baseline controls (average zip code-level income, the fraction of IRS filings under \$25,000, between \$25,000-50,000, \$50,000-75,000, \$75,000-100,000, and over \$100,000, credit score quintiles, median credit score, total delinquent debt balances, unemployment rate, median house value, the fraction of the population with a college education, the divorce rate, and population density), county fixed effects, and year fixed effects are included in all specifications.  All regressions are weighted by population. Robust standard errors clustered at the county-year level are in parentheses. Effective number of clusters: 36.0. Wild Cluster Bootstrap p-values are reported for our main parameter of interest. \sym{*} \(p<0.10\), \sym{**} \(p<0.05\), \sym{***} \(p<0.01\). 
\end{tablenotes}
\end{threeparttable}
\end{table}

%% file: Tables/Tables_share/results_judgment_type_neighborhood_share_black.tex
\begin{sidewaystable}[htbp]\centering
\begin{threeparttable}
\footnotesize
\def\sym#1{\ifmmode^{#1}\else\(^{#1}\)\fi}
\caption{Attorney Representation and Judgment Type: Share Black\label{tab:judgment_type_share_black}}
\begin{tabular}{l*{8}{c}}
\hline\hline
                    &\multicolumn{1}{c}{(1)}&\multicolumn{1}{c}{(2)}&\multicolumn{1}{c}{(3)}&\multicolumn{1}{c}{(4)}&\multicolumn{1}{c}{(5)}&\multicolumn{1}{c}{(6)}&\multicolumn{1}{c}{(7)}&\multicolumn{1}{c}{(8)}\\
                    &\multicolumn{1}{c}{Attorney}&\multicolumn{1}{c}{Judgments}&\multicolumn{1}{c}{Consent}&\multicolumn{1}{c}{Constested}&\multicolumn{1}{c}{Default}&\multicolumn{1}{c}{Dismissed}&\multicolumn{1}{c}{Settle}&\multicolumn{1}{c}{Judgments}\\
\hline
&&&&& & & & \\[\dimexpr-\normalbaselineskip+2pt]
Share Black: ZIP    &      -0.036\sym{***}&       1.409\sym{***}&       1.474\sym{***}&       0.014         &       0.011\sym{*}  &       0.045\sym{***}&       0.010         &      1.575\sym{***}\\
                    &     (0.006)         &     (0.282)         &     (0.283)         &     (0.011)         &     (0.006)         &     (0.009)         &     (0.011)         &     (0.289)         \\
\hline
&&&&& & & & \\[\dimexpr-\normalbaselineskip+2pt]
Attorney Representation & & X & X&X&X&X&X&X \\
Lagged Case Outcomes & & & & & & & & X \\ \hline 
&&&&& & & & \\[\dimexpr-\normalbaselineskip+2pt]
Wild Cluster Bootstrap p-value&           0         &       .0002         &           0         &       .2137         &       .0987         &           0         &       .3535         &           0.0001         \\
Observations        &    7019      &    7019      &    7019      &    7019      &    7019      &    7019      &    7019      &    6093     \\
$R^2$               &       0.450         &       0.710         &       0.697         &       0.470         &       0.438         &       0.397         &       0.650         &       0.730         \\
\hline\hline
\end{tabular}
\begin{tablenotes}
\scriptsize
\item Notes:  This table explores the extent to which attorney representation or previous case outcomes can explain the racial gap in judgments, using the share of black residents in a zip code as our measure for racial composition.  The sample includes Missouri zip codes in the common support sample. The outcome variable in Column (1) is the share of debt collection court cases where an attorney represented the defendant.  The outcome variable in columns (2) and (8) are judgments per 100 people.  The outcome variables in Columns (3)-(7) are the share of cases resulting in a consent judgment, a contested judgment, a default judgment, the case being dismissed, and the case being settled, respectively.  All regressions are weighted by population. Baseline controls (average zip code-level income, the fraction of IRS filings under \$25,000, between \$25,000-50,000, \$50,000-75,000, \$75,000-100,000, and over \$100,000, credit score quintiles, median credit score, total delinquent debt balances, unemployment rate, median house value, the fraction of the population with a college education, the divorce rate, and population density), county fixed effects, and year fixed effects are included in all specifications. Aside from Column (1), all specifications control for the share of cases with attorney representation. Column (8) controls for the share of default judgments. Robust standard errors clustered at the county-year level are in parentheses. Effective number of clusters: 28.0. Wild Cluster Bootstrap p-values are reported for our main parameter of interest. \sym{*} \(p<0.10\), \sym{**} \(p<0.05\), \sym{***} \(p<0.01\). 
\end{tablenotes}
\end{threeparttable}
\end{sidewaystable}

%% file: Tables/Tables_neighborhood/results_debt_portfolio_neighborhood_bisg.tex
\begin{table}[htbp]\centering
\begin{threeparttable}
\footnotesize
\def\sym#1{\ifmmode^{#1}\else\(^{#1}\)\fi}
\caption{Debt Portfolios\label{tab:debt_composition_bisg}}
\begin{tabular}{l*{5}{c}}
\hline\hline
                    &\multicolumn{1}{c}{(1)}&\multicolumn{1}{c}{(2)}&\multicolumn{1}{c}{(3)}&\multicolumn{1}{c}{(4)}&\multicolumn{1}{c}{(5)}\\
                       &\multicolumn{5}{c}{Judgment Rate}\\
\hline
Black Majority: BISG&      0.6222\sym{***}&      0.6329\sym{***}&      0.6264\sym{***}&      0.6302\sym{***}&      0.6266\sym{***}\\
                    &    (0.1308)         &    (0.1349)         &    (0.1343)         &    (0.1274)         &    (0.1229)         \\
\hline
Debt Levels         &           X         &                     &                     &                     &           X         \\
Monthly Payment and Utilization&                     &           X         &                     &                     &           X         \\
Debt Composition    &                     &                     &           X         &                     &           X         \\
Delinquency/Bankruptcy/Collections&                     &                     &                     &           X         &           X         \\
Wild Cluster Bootstrap p-value&       .0004         &       .0004         &       .0005         &       .0002         &       .0002         \\
Observations        &   7015       &   7015        &   7015      &   7015        &   7015        \\
$R^2$               &      0.6893         &      0.6869         &      0.6861         &      0.6884         &      0.6940         \\
\hline\hline
\end{tabular}
\begin{tablenotes}
\scriptsize
\item  This table presents results of a regression of the judgment rate on neighborhood racial composition (as defined using the BISG algorithm) and other neighborhood-level characteristics.  The sample includes Missouri zip codes in the common support sample.  The dependent variable is judgments per 100 individuals.  Baseline controls include average zip code-level income, the fraction of IRS filings under \$25,000, between \$25,000-50,000, \$50,000-75,000, \$75,000-100,000, and over \$100,000, credit score quintiles, median credit score, total delinquent debt balances, unemployment rate, median house value, the fraction of the population with a college education, the divorce rate, and population density.  Baseline controls, county fixed effects, and year fixed effects are included in all specifications, and all regressions are weighted by population. Robust standard errors clustered at the county-year level are in parentheses. Effective number of clusters: 36.0. Wild Cluster Bootstrap p-values are reported for our main parameter of interest. \sym{*} \(p<0.10\), \sym{**} \(p<0.05\), \sym{***} \(p<0.01\). 
\end{tablenotes}
\end{threeparttable}
\end{table}

%% file: Tables/Tables_neighborhood/summary_tab1_other_areas.tex
\begin{table}[b]
\centering
\begin{threeparttable}
\footnotesize
\def\sym#1{\ifmmode^{#1}\else\(^{#1}\)\fi}
\caption{Summary Statistics: IL \& NJ\label{tab:summary_statistics_il_nj}}
\begin{tabular}{l*{3}{cc}}
\hline\hline
                    &\multicolumn{1}{c}{Black}&\multicolumn{1}{c}{Non-Black}&\multicolumn{1}{c}{Difference}\\
\hline
& & & \\[\dimexpr-\normalbaselineskip+2pt]
\textbf{Panel A: Judgments and Financial Characteristics} & & & \\
 \textcolor{white}{...}Judgments per 100 People &        2.70&        1.72&       -0.98\sym{***}\\
                    &      (2.00)&      (1.79)&                     \\
 \textcolor{white}{...}Median Credit Score  &      588.98&      707.15&      118.17\sym{***}\\
                    &     (36.72)&     (54.62)&                     \\
 \textcolor{white}{...}90+ DPD Debt Balances &     7135.13&     4260.03&    -2875.10\sym{***}\\
                    &   (6609.97)&   (6448.52)&                     \\
\textbf{Panel B: Additional Neighborhood Characteristics } & & & \\ 
 \textcolor{white}{...}GINI Index       &        0.46&        0.41&       -0.06\sym{***}\\
                    &      (0.05)&      (0.05)&                     \\
 \textcolor{white}{...}Unemployment Rate  &        0.12&        0.07&       -0.06\sym{***}\\
                    &      (0.03)&      (0.02)&                     \\
 \textcolor{white}{...}Divorce Rate  &        0.11&        0.10&       -0.02\sym{***}\\
                    &      (0.02)&      (0.03)&                     \\
 \textcolor{white}{...}Fraction with Bachelors Degree&        0.18&        0.30&        0.11\sym{***}\\
                    &      (0.10)&      (0.14)&                     \\
 \textcolor{white}{...}Gross Rent: Median   &        0.97&        1.06&        0.09\sym{***}\\
                    &      (0.25)&      (0.29)&                     \\
 \textcolor{white}{...}Home Ownership Rate      &        0.51&        0.71&        0.20\sym{***}\\
                    &      (0.20)&      (0.17)&                     \\
 \textcolor{white}{...}Banks (5 miles)  &      166.92&      113.30&      -53.63\sym{***}\\
                    &    (138.48)&    (170.14)&                     \\
 \textcolor{white}{...}Payday Lenders (5 miles)&       33.65&       11.84&      -21.81\sym{***}\\
                    &     (39.86)&     (25.73)&                     \\
 \textcolor{white}{...}Zillow: Single Family House&        176.37&      189.89&       13.53         \\
                    &     (92.22)&    (116.21)&                     \\
 \textcolor{white}{...}HMDA: Share Rejected, Black - White  &      175.91&      252.34&       76.43\sym{***}\\
                    &     (90.21)&    (123.72)&                     \\
 \textcolor{white}{...}Mean Household Income (IRS) &        0.16&        0.13&       -0.03\sym{***}\\
                    &      (0.11)&      (0.19)&                     \\
\hline
Observations        &         200&         705&         905         \\
\hline\hline
\end{tabular}
\begin{tablenotes}
        \scriptsize
        \item Notes:  This table presents summary statistics for the Illinois and New Jersey neighborhood level data from 2004-2013.  The sample includes only observations on the common support.  Column 1 displays the sample means and standard deviations for neighborhoods in which at least 50\% of the population is black and column 2 displays sample means and standard deviations for neighborhoods in which less than 50\% is black.  The third column reports the difference in means across the two samples. \sym{*} \(p<0.10\), \sym{**} \(p<0.05\), \sym{***} \(p<0.01\)
\end{tablenotes}
\end{threeparttable}
\end{table}

%% file: Tables/Tables_neighborhood/results_debt_portfolio_neighborhood_other_areas.tex
\begin{table}[htbp]\centering
\begin{threeparttable}
\footnotesize
\def\sym#1{\ifmmode^{#1}\else\(^{#1}\)\fi}
\caption{Debt Portfolios (NJ and IL Data)\label{tab:debt_composition_nj_il}}
\begin{tabular}{l*{5}{c}}
\hline\hline
                    &\multicolumn{1}{c}{(1)}&\multicolumn{1}{c}{(2)}&\multicolumn{1}{c}{(3)}&\multicolumn{1}{c}{(4)}&\multicolumn{1}{c}{(5)}\\
                       &\multicolumn{5}{c}{Judgment Rate}\\
  
\hline
Black Majority:  ZIP     &      0.2748\sym{***}&      0.2753\sym{***}&      0.2876\sym{***}&      0.2840\sym{***}&      0.2592\sym{***}\\
                    &    (0.0858)         &    (0.0850)         &    (0.0871)         &    (0.0924)         &    (0.0870)         \\
\hline
Wild Cluster Bootstrap p-value&       .0012         &       .0011         &       .0012         &       .0018         &       .0033         \\
Debt Levels         &           X         &                     &                     &                     &           X         \\
Monthly Payment and Utilization&                     &           X         &                     &                     &           X         \\
Debt Composition    &                     &                     &           X         &                     &           X         \\
Delinquency/Bankruptcy/Collections&                     &                     &                     &           X         &           X         \\
Observations        &   3103        &   3103        &   3103        &   3103        &   3103        \\
$R^2$               &      0.9105         &      0.9113         &      0.9101         &      0.9110         &      0.9145         \\
\hline\hline
\end{tabular}
\begin{tablenotes}
\scriptsize
\item Notes:  This table presents results of a regression of the judgment rate on neighborhood racial composition and other neighborhood-level characteristics.  The sample includes New Jersey and Illinois zip codes in the common support sample.  The dependent variable is judgments per 100 individuals.  Baseline controls include average zip code-level income, the fraction of IRS filings under \$25,000, between \$25,000-50,000, \$50,000-75,000, \$75,000-100,000, and over \$100,000, credit score quintiles, median credit score, total delinquent debt balances, unemployment rate, median house value, the fraction of the population with a college education, the divorce rate, and population density.  Baseline controls, county fixed effects, and year fixed effects are included in all specifications, and all regressions are weighted by population. Robust standard errors clustered at the county-year level are in parentheses. Effective number of clusters: 36.0. Wild Cluster Bootstrap p-values are reported for our main parameter of interest. \sym{*} \(p<0.10\), \sym{**} \(p<0.05\), \sym{***} \(p<0.01\). 
\end{tablenotes}
\end{threeparttable}
\end{table}

%% file: Tables/Tables_noncommon/summary_tab1_off_common_support.tex
\begin{table}[b]\centering
  \begin{threeparttable}
 \footnotesize
\def\sym#1{\ifmmode^{#1}\else\(^{#1}\)\fi}
\caption{Summary Statistics: Entire Sample\label{tab:summary_statistics_off_common}}
\begin{tabular}{l*{3}{cc}}
\hline\hline
                    &\multicolumn{1}{c}{Black}&\multicolumn{1}{c}{Non-Black}&\multicolumn{1}{c}{Difference}\\
\hline
& & & \\[\dimexpr-\normalbaselineskip+2pt]
\textbf{Panel A: Judgments and Finacial Characteristics} & & & \\
 \textcolor{white}{...}Judgments per 100 People      &        2.65&        1.32&       -1.32\sym{***}\\
                    &      (1.35)&      (1.53)&                     \\
 \textcolor{white}{...}Share of Default Judgments     &        0.45&        0.37&       -0.08\sym{***}\\
                    &      (0.07)&      (0.17)&                     \\
 \textcolor{white}{...}Share of Consent Judgments          &        0.16&        0.17&        0.02\sym{***}\\
                    &      (0.07)&      (0.14)&                     \\
 \textcolor{white}{...}Share of Dismissed Judgments    &        0.12&        0.15&        0.03\sym{***}\\
                    &      (0.11)&      (0.15)&                     \\
 \textcolor{white}{...}Share of  Constested Judgments        &        0.05&        0.05&       -0.01\sym{*}  \\
                    &      (0.05)&      (0.09)&                     \\
 \textcolor{white}{...}Share of  Settled Cases        &        0.19&        0.21&        0.02\sym{***}\\
                    &      (0.09)&      (0.17)&                     \\
 \textcolor{white}{...}Share w/ Attorney   &        0.04&        0.10&        0.06\sym{***}\\
                    &      (0.02)&      (0.11)&                     \\
 \textcolor{white}{...}Median Credit Score &      586.65&      688.71&      102.06\sym{***}\\
                    &     (50.98)&     (76.50)&                     \\
 \textcolor{white}{...}90+ DPD Debt Balances       &     3193.18&     1416.45&    -1776.73\sym{***}\\
                    &   (2676.72)&   (4253.49)&                     \\
\textbf{Panel B: Additional Neighborhood Characteristics} & & & \\ 
 \textcolor{white}{...}GINI Index &        0.46&        0.40&       -0.06\sym{***}\\
                    &      (0.05)&      (0.07)&                     \\

 \textcolor{white}{...}Unemployment Rate  &        0.12&        0.05&       -0.07\sym{***}\\
                    &      (0.05)&      (0.04)&                     \\
 \textcolor{white}{...}Divorce Rate&        0.13&        0.12&       -0.01\sym{***}\\
                    &      (0.03)&      (0.06)&                     \\
 \textcolor{white}{...}Fraction with Bachelors Degree&              0.16&        0.17&        0.00         \\
                    &      (0.10)&      (0.13)&                     \\
\textcolor{white}{...}Home Ownership Rate &        0.47&        0.76&        0.29\sym{***}\\
                    &      (0.19)&      (0.14)&                     \\
\textcolor{white}{...}Banks (5 miles)     &       85.44&       11.88&      -73.56\sym{***}\\
                    &     (40.79)&     (30.95)&                     \\
 \textcolor{white}{...}Payday Lenders (5 miles)&       29.98&        3.23&      -26.75\sym{***}\\
                    &      (9.86)&      (7.76)&                     \\
 \textcolor{white}{...}Zillow: Single Family House (in 000s) &       90.15&      122.26&       32.11\sym{***}\\
                    &     (54.29)&     (68.87)&                     \\
 \textcolor{white}{...}HMDA: Share Rejected, Black - White&        0.19&        0.15&       -0.04\sym{***}\\
                    &      (0.11)&      (0.31)&                     \\
 \textcolor{white}{...}Mean Household Income from IRS (in 000s) &       28.67&       41.01&       12.34\sym{***}\\
                    &     (10.00)&     (24.25)&                     \\

\hline
Zip Codes       &         258&        9530&           \\
Population &    17697 &     5808 &   \\
\hline\hline
\end{tabular}
   \begin{tablenotes}
     \scriptsize
    \item Notes:  This table presents summary statistics for the entire Missouri neighborhood level data from 2004-2013. Column 1 displays the sample means and standard deviations for neighborhoods in which at least 50\% of the population is black and column 2 displays sample means and standard deviations for neighborhoods in which less than 50\% is black.  The third column reports the difference in means across the two samples. \sym{*} \(p<0.10\), \sym{**} \(p<0.05\), \sym{***} \(p<0.01\)
   \end{tablenotes}
\end{threeparttable}
\end{table}

%% file: Tables/Tables_noncommon/results_debt_portfolio_neighborhood_off_common_support.tex
\begin{table}[htbp] \centering
\begin{threeparttable}
\footnotesize
\def\sym#1{\ifmmode^{#1}\else\(^{#1}\)\fi}
\caption{Judgments and Debt Portfolios: Entire Sample\label{tab:debt_composition_off_common}}
\begin{tabular}{l*{5}{c}}
\hline\hline
                    &\multicolumn{1}{c}{(1)}&\multicolumn{1}{c}{(2)}&\multicolumn{1}{c}{(3)}&\multicolumn{1}{c}{(4)}&\multicolumn{1}{c}{(5)}\\
                       &\multicolumn{5}{c}{Judgment Rate}\\
\hline
& & & & & \\[\dimexpr-\normalbaselineskip+2pt]
Black Majority: ZIP &      0.6926\sym{***}&      0.7086\sym{***}&      0.7058\sym{***}&      0.6957\sym{***}&      0.6871\sym{***}\\
                    &    (0.1549)         &    (0.1599)         &    (0.1594)         &    (0.1558)         &    (0.1476)         \\
\hline
& & & & &  \\[\dimexpr-\normalbaselineskip+2pt]
Debt Levels         &           X         &                     &                     &                     &           X         \\
Monthly Payment and Utilization&                     &           X         &                     &                     &           X         \\
Debt Composition    &                     &                     &           X         &                     &           X         \\
Delinquency/Bankruptcy/Collections&                     &                     &                     &           X         &           X         \\
\hline
& & & & & \\[\dimexpr-\normalbaselineskip+2pt]
Wild Cluster Bootstrap p-value&       .0004         &       .0005         &       .0005         &       .0004         &       .0003         \\
Observations        &   7484        &   7484        &   7484        &   7484        &   7484        \\
$R^2$               &      0.7037         &      0.7021         &      0.7020         &      0.7038         &      0.7079         \\
\hline\hline
\end{tabular}
\begin{tablenotes}
\scriptsize
\item Notes:  This table presents results of a regression of the judgment rate on neighborhood racial composition and other neighborhood-level characteristics.  The sample includes the entire Missouri zip code sample.  The dependent variable is judgments per 100 individuals.  Baseline controls include average zip code-level income, the fraction of IRS filings under \$25,000, between \$25,000-50,000, \$50,000-75,000, \$75,000-100,000, and over \$100,000, credit score quintiles, median credit score, total delinquent debt balances, unemployment rate, median house value, the fraction of the population with a college education, the divorce rate, and population density.  Baseline controls, county fixed effects, and year fixed effects are included in all specifications, and all regressions are weighted by population. Robust standard errors clustered at the county-year level are in parentheses. Effective number of clusters: 36.0. Wild Cluster Bootstrap p-values are reported for our main parameter of interest. \sym{*} \(p<0.10\), \sym{**} \(p<0.05\), \sym{***} \(p<0.01\). 
\end{tablenotes}
\end{threeparttable}
\end{table}

%% file: Tables/Tables_noncommon/results_judgment_type_neighborhood_off_common_support.tex
\begin{sidewaystable}[htbp]\centering
\begin{threeparttable}
\footnotesize
\def\sym#1{\ifmmode^{#1}\else\(^{#1}\)\fi}
\caption{Attorney Representation and Judgment Type: Entire Sample\label{tab:judgment_type_off_common}}
\begin{tabular}{l*{9}{c}}
\hline\hline
                    &\multicolumn{1}{c}{(1)}&\multicolumn{1}{c}{(2)}&\multicolumn{1}{c}{(3)}&\multicolumn{1}{c}{(4)}&\multicolumn{1}{c}{(5)}&\multicolumn{1}{c}{(6)}&\multicolumn{1}{c}{(7)}&\multicolumn{1}{c}{(8)}&\multicolumn{1}{c}{(9)}\\
                    &\multicolumn{1}{c}{Attorney}&\multicolumn{1}{c}{Judgments}&\multicolumn{1}{c}{Judgments}&\multicolumn{1}{c}{Consent}&\multicolumn{1}{c}{Contested}&\multicolumn{1}{c}{Default}&\multicolumn{1}{c}{Dismissed}&\multicolumn{1}{c}{Settle}&\multicolumn{1}{c}{Judgments}\\
\hline
&&&&& & & & \\[\dimexpr-\normalbaselineskip+2pt]
Black Majority: ZIP &      -0.016\sym{***}&       0.685\sym{***}&       0.706\sym{***}&       0.004         &       0.001         &       0.014\sym{***}&       0.010         &      -0.022\sym{***} &       0.752\sym{***}\\
                    &     (0.003)         &     (0.155)         &     (0.160)         &     (0.005)         &     (0.004)         &     (0.005)         &     (0.007)         &     (0.006)    & (0.162)     \\
\hline
&&&&& & & & &\\[\dimexpr-\normalbaselineskip+2pt]
Attorney Representation & & X & X&X&X&X&X&X&X \\
Lagged Case Outcomes & & & & & & & & & X \\ \hline 
&&&&& & & & S \\[\dimexpr-\normalbaselineskip+2pt]
Wild Cluster Bootstrap p-value&           0         &       .0004         &       .0004         &       .4475         &       .8498         &       .0096         &       .1681         &       .0036    & 0.0008     \\
Observations        &    7484       &    7484       &    7484       &    7484       &    7484       &    7484       &    7484       &    7484    & 6701   \\
$R^2$               &       0.440         &       0.715         &       0.702         &       0.479         &       0.407         &       0.389         &       0.654         &       0.643     & 0.732    \\
\hline\hline
\end{tabular}
\begin{tablenotes}
\scriptsize
\item Notes:  This table explores the extent to which attorney representation or previous case outcomes can explain the racial gap in judgments.  The sample includes the entire Missouri zip code level sample. The outcome variable in Column (1) is the share of debt collection court cases where an attorney represented the defendant.  The outcome variable in columns (2) and (8) are judgments per 100 people.  The outcome variables in Columns (3)-(7) are the share of cases resulting in a consent judgment, a contested judgment, a default judgment, the case being dismissed, and the case being settled, respectively.  All regressions are weighted by population. Baseline controls (average zip code-level income, the fraction of IRS filings under \$25,000, between \$25,000-50,000, \$50,000-75,000, \$75,000-100,000, and over \$100,000, credit score quintiles, median credit score, total delinquent debt balances, unemployment rate, median house value, the fraction of the population with a college education, the divorce rate, and population density), county fixed effects, and year fixed effects are included in all specifications. Aside from Column (1), all specifications control for the share of cases with attorney representation. Column (8) controls for the share of default judgments. Robust standard errors clustered at the county-year level are in parentheses. Effective number of clusters: 36.0. Wild Cluster Bootstrap p-values are reported for our main parameter of interest. \sym{*} \(p<0.10\), \sym{**} \(p<0.05\), \sym{***} \(p<0.01\). 
\end{tablenotes}
\end{threeparttable}
\end{sidewaystable}

%% file: Tables/Tables_neighborhood/results_other_races_neighborhood.tex
\begin{table}[htbp] \centering
\begin{threeparttable}
\footnotesize
\def\sym#1{\ifmmode^{#1}\else\(^{#1}\)\fi}
\caption{Judgments and Other Demographic Groups\label{tab:race_other}}
\begin{tabular}{l*{6}{c}}
\hline\hline
                    &\multicolumn{1}{c}{(1)}&\multicolumn{1}{c}{(2)}&\multicolumn{1}{c}{(3)}&\multicolumn{1}{c}{(4)}&\multicolumn{1}{c}{(5)}&\multicolumn{1}{c}{(6)}\\
                       &\multicolumn{6}{c}{Judgment Rate}\\
\hline
& & & & & & \\[\dimexpr-\normalbaselineskip+2pt]
Share Black: ZIP    &      2.2881\sym{***}&      1.5574\sym{***}&      1.9195\sym{***}&      1.4544\sym{***}&      1.4865\sym{***}&      1.6260\sym{***}\\
                    &    (0.2748)         &    (0.3184)         &    (0.2505)         &    (0.2955)         &    (0.2777)         &    (0.0610)         \\
& & & & & & \\[\dimexpr-\normalbaselineskip+2pt]
Share Asian: ZIP    &     -5.6745\sym{***}&     -3.3050\sym{***}&     -4.8311\sym{***}&     -3.1980\sym{***}&     -1.6851\sym{**} &     -3.6100\sym{***}\\
                    &    (0.9938)         &    (0.9074)         &    (0.9953)         &    (0.8989)         &    (0.7889)         &    (0.4954)         \\
& & & & & & \\[\dimexpr-\normalbaselineskip+2pt]
Share Hispanic: ZIP &      1.0901\sym{***}&     -0.3160\sym{***}&      0.7415\sym{***}&     -0.3814\sym{***}&      0.5452\sym{***}&     -0.3274\sym{*}  \\
                    &    (0.1345)         &    (0.1217)         &    (0.1357)         &    (0.1297)         &    (0.1421)         &    (0.1820)         \\
\hline
& & & & & & \\[\dimexpr-\normalbaselineskip+2pt]
Baseline Controls   &                     &                     &                     &                     &           X         &                     \\
Income     &                     &           X         &                     &           X         &           X         &                     \\
Credit Score  &                     &                     &           X         &           X         &           X         &                     \\
Lagged Baseline Controls&                     &                     &                     &                     &                     &           X         \\ \hline
& & & & & & \\[\dimexpr-\normalbaselineskip+2pt]
Wild Cluster Bootstrap p$_{\text{Black}}$&           0         &           0         &           0         &           0         &           0         &           0         \\
Wild Cluster Bootstrap p$_{\text{Hispanic}}$&           0         &       .0137         &       .0001         &        .008         &       .0003         &       .1084         \\
Wild Cluster Bootstrap p$_{\text{Asian}}$&           0         &       .0031         &       .0002         &       .0026         &       .0552         &           0         \\
Observations        &   7365        &   7365        &   7019        &   7019        &   7019        &   6105        \\
$R^2$               &      0.6496         &      0.6796         &      0.6618         &      0.6841         &      0.6973         &      0.6952         \\
\hline\hline
\end{tabular}
\begin{tablenotes}
\scriptsize
\item Notes:  This table presents results of a regression of the judgment rate on neighborhood racial composition and other neighborhood-level characteristics.  The sample includes Missouri zip codes in the common support sample.  The dependent variable is judgments per 100 individuals.  Baseline controls include average zip code-level income, the fraction of IRS filings under \$25,000, between \$25,000-50,000, \$50,000-75,000, \$75,000-100,000, and over \$100,000, credit score quintiles, median credit score, total delinquent debt balances, unemployment rate, median house value, the fraction of the population with a college education, the divorce rate, and population density.  Baseline controls, county fixed effects, and year fixed effects are included in all specifications, and all regressions are weighted by population. Robust standard errors clustered at the county-year level are in parentheses. Effective number of clusters: 36.0. Wild Cluster Bootstrap p-values are reported for our main parameter of interest. \sym{*} \(p<0.10\), \sym{**} \(p<0.05\), \sym{***} \(p<0.01\). 
\end{tablenotes}
\end{threeparttable}
\end{table}

%% file: Tables/Tables_neighborhood/dml.tex
\begin{table}[htbp]\centering
\begin{threeparttable}
\footnotesize
\caption{Judgments and Double Machine Learning\label{tab:dml}}
\begin{tabular}{l*{5}{c}} \hline \hline
 & R2: Y Model & R2: T Model & Estimates & T-values \\ \hline 
 & & &  & \\[\dimexpr-\normalbaselineskip+2pt]
T: Majority Black & 0.5121 & 0.9877 & 1.020 & 1.824 \\
 & [0.0069] & [0.0022] & [0.2682] & [0.4028] \\ \hline \hline
% & & &  & \\[\dimexpr-\normalbaselineskip+2pt]
%T: Continuous & 0.5121 &  0.9904 & 3.909 & 6.8267 \\
% & [0.0069] & [0.0015] & [0.4592] & [1.3375] \\ \hline \hline 
\end{tabular}
\begin{tablenotes}
\scriptsize
\item Notes: We use five-fold cross-validation in the first stage. Features used correspond to Column (5) of Table \ref{tab:creditworthiness}. For the classification tasks, we use Gradient Boosted Trees, while for the regression tasks, we use Random Forest.  We repeat the estimation with different seeds 1,000 times with distinct train-test splits and the statistics reported are the averages obtained from this exercise. Standard deviations are in brackets.
\end{tablenotes}
\end{threeparttable}
\end{table}